\documentclass[aps,prd,showpacs,superscriptaddress,preprintnumbers,dblfloatfix,floatfix,nofootinbib,twocolumn,12points]{revtex4-2}
\usepackage{graphicx} 

\usepackage{amsmath,amssymb}
\usepackage{epsfig}
\usepackage{hyperref}
\usepackage{color}
\usepackage{slashed}
\usepackage{placeins}
\usepackage{ulem}
\usepackage{amsfonts}
\usepackage{graphicx, rotating}
\usepackage{epstopdf}
\usepackage{latexsym}
\usepackage{rotating}
\usepackage{braket}
\usepackage{dsfont}
\usepackage[dvipsnames]{xcolor}

\usepackage[export]{adjustbox}
\usepackage{multirow}

\usepackage[font={small}]{caption}   

\definecolor{myred}{rgb}{0.7, 0, 0}
\definecolor{myblue}{rgb}{0, 0, 0.7}
\definecolor{mygreen}{rgb}{0.04, 0.7, 0.5}
\definecolor{mygray}{rgb}{0.1, 0.1, 0.1}

\hypersetup{colorlinks,citecolor=myred,linkcolor=myblue,urlcolor=myblue,linktocpage=true}

 \def\be   {\begin{equation}}   \def\ee   {\end{equation}}
 \def\ba   {\begin{array}}      \def\ea   {\end{array}}
 \def\bea  {\begin{eqnarray}}   \def\eea  {\end{eqnarray}}
 \def\bean {\begin{eqnarray*}}  \def\eean {\end{eqnarray*}}
 \def\nn{\nonumber}
 \def\bry{\begin{array}}
 \def\ery{\end{array}}
\def\TeV{\,{\rm TeV}}
\def\GeV{\,{\rm GeV}}
\def\MeV{\,{\rm MeV}}

\def\eV{\,{\rm eV}}
\def\Mpl{ M_{\rm Pl}}
\def\vew{v_{\rm EW}}

\def\tin{t_{\rm in}}

\def\tosc{t_{\rm osc}}
\def\ain{R_{\rm in}}
\def\arh{R_{\rm rh}}
\def\af{R_{f}}
\def\abeg{R_{\rm beg}}
\def\aend{R_{\rm end}}

\def\Hrh{H_{\rm rh}}
\def\aphi{R_\phi}
\def\ascan{R_{\rm scan}}
\def\Trh{T_{\rm rh}}
\def\TSM{T_{\gamma}}
\def\mB{m_{\rm B}}
\def\nB{n_{\rm B}}
\def\mDM{m_{\rm DM}}
\def\nDM{n_{\rm DM}}
\def\na{n_{a}}
\def\fphi{f_{\phi}}
\def\rhoSM{\rho_{\gamma}}
\def\tbeg{t_{\rm beg}}
\def\tend{t_{\rm end}}
\def\tin{t_{\rm in}}
\def\trh{t_{\rm rh}}
\def\tf{t_{f}}
\def\tphi{t_{\phi}}
\def\l{\left(}
\def\r{\right)}

\newcommand{\skipnew}[1]{}

\numberwithin{equation}{section}
\renewcommand{\theequation}{\arabic{section}.\arabic{equation}}

\def\UMD{\small{Maryland Center for Fundamental Physics, University of Maryland, College Park, MD 20742, USA}}

\def\Michigan{\small{Leinweber Center for Theoretical Physics, Department of Physics, University of Michigan, Ann Arbor, MI 48109, USA}}

\usepackage{tikz,xcolor,hyperref}
\definecolor{lime}{HTML}{A6CE39}
\DeclareRobustCommand{\orcidicon}{%
	\begin{tikzpicture}
	\draw[lime, fill=lime] (0,0) 
	circle [radius=0.16] 
	node[white] {{\fontfamily{qag}\selectfont \tiny ID}};	\draw[white, fill=white] (-0.0625,0.095) 
	circle [radius=0.007];	\end{tikzpicture}
	\hspace{-2mm}}
\foreach \x in {A,...,Z}{%
	\expandafter\xdef\csname orcid\x\endcsname{\noexpand\href{https://orcid.org/\csname orcidauthor\x\endcsname}{\noexpand\orcidicon}}
}

\date{\today}

\begin{document}
\title{Predicting the Dark Matter - Baryon Abundance Ratio}
\author{Abhishek Banerjee\orcidA{}\,}
\email{abanerj4@umd.edu}
\affiliation{\UMD}
\author{Dawid Brzeminski\orcidB{}\, }
\email{dawid@umich.edu}
\affiliation{\UMD}
\affiliation{\Michigan}
\author{Anson Hook\orcidC{}\,}
\email{hook@umd.edu}
\affiliation{\UMD}

\begin{abstract}

We discuss relaxation solutions to the dark matter - baryon coincidence problem in the context of QCD axion dark matter. 
In relaxation solutions, a moduli dynamically adjusts the mass of dark matter and baryons until their energy densities are $\mathcal{O}(1)$ the same.  Because the QCD axion is heavily connected to QCD, scanning the QCD axion mass inherently also scans the proton mass.  In the context of relaxation solutions, this implies that the ratio of dark matter to baryon abundances ($\Omega_{\rm DM}/\Omega_{\rm B}$) is a ratio of beta functions showing that these models can only accommodate discrete values of $\Omega_{\rm DM}/\Omega_{\rm B}$ thereby ``predicting" the ratio of the dark matter to baryon abundances.  The original composite axion model has only a single integer degree of freedom $N$, the size of the gauge group, and we show that when $N=8$ the observed value of $\Omega_{\rm DM}/\Omega_{\rm B} = 5.36$ is reproduced to within its percent level error bars.  Novel tests of this model include more precise measurements of $\Omega_{\rm DM}/\Omega_{\rm B}$, a better lattice determination of the dependence of the proton mass on the high energy QCD gauge coupling, as well as more traditional tests such as fifth force experiments.
\end{abstract}

\maketitle
\preprint{LCTP-24-20}

\section{Introduction}

There is a remarkable cosmological coincidence between the relic cold dark matter (DM) energy density ($\Omega_{\rm DM}$) and the relic baryon energy density ($\Omega_{\rm B}$). 
The current measured value of the ratio is 
\bea
r_{\rm obs} \equiv \frac{\Omega_{\rm DM}}{\Omega_{\rm B}} = 5.36\pm 0.05\,,
\eea
where each of these quantities are measured at a sub-percent level~\cite{Planck:2018vyg}. 
While this ratio is technically an $\mathcal{O}(1)$ number, its proximity to unity is non-trivial. The baryon and DM abundances are set by completely different mechanisms~\footnote{In contrast to other $\mathcal{O}(1)$ coincidences, such as $\Omega_{\rm DE}/\Omega_{\rm DM} \sim 3$, the ratio $\Omega_{\rm DM}/\Omega_{\rm B}$ remains constant.}.

The QCD confinement scale  $\Lambda_{\rm QCD}$ is generated by dimensional transmutation and is dictated by a theory parameter that at high energy (UV) happens to be $\sim 0.1$. 
Dimensional transmutation makes the baryon mass, and thus $\Omega_{\rm B}$, exponentially sensitive to the $\mathcal{O}(1)$ UV theory parameter and is the explanation for why the proton mass is eighteen orders of magnitude lighter than the Planck scale. 
On the other hand, the DM mass and number density depends on its theory and production mechanism, each possessing its own strong sensitivity to their own theory parameters, which are largely unrelated to QCD. 
All of these unrelated parameters conspiring to produce a result that is only a factor of $5.36$ different makes this puzzle astonishing.

Several solutions have been proposed to address this coincidence ~\cite{Hodges:1993yb,Berezhiani:1995am,Nussinov:1985xr,Gelmini:1986zz,Barr:1990ca,Barr:1991qn,Kaplan:1991ah,Kaplan:2009ag,Foot:2003jt,An:2009vq,Farina:2015uea,GarciaGarcia:2015pnn,Lonsdale:2018xwd,Ibe:2019ena,Murgui:2021eqf,Bai:2013xga,Newstead:2014jva,Ritter:2022opo,Rosa:2022sym}.  Many of these models, such as asymmetric DM and mirror world scenarios, introduce relations between either the baryon and DM number densities or between their masses. For instance, in mirror models the mirror proton mass can naturally differ from the SM proton mass by an $\mathcal{O}(1)$ factor, giving $r_{\rm obs}$ of the right order. In contrast, our approach is based on a relaxation mechanism where the ratio of relic densities is dynamically selected by the cosmological evolution of a scanner~\cite{Brzeminski:2023wza}. This provides a qualitatively different resolution of the coincidence.

To see how it works,  let us consider a scenario where both the baryon mass ($\mB$) and the DM mass ($m_{\rm DM}$) depend on the expectation value of a scalar field, $\phi$, as
\bea
\!\!\!\!\!\!\!\!\!\!
&\mB& \to \mB(\phi) = \mB(0)\exp{\left(c_B\phi/f_\phi\right)}\nn\\
&\mDM& \to \mDM(\phi) = \mDM(0)\exp{\left(c_D\phi/f_\phi\right)}\,,
\label{Eq:baryon_DM_phi}
\eea
where $c_B$ and $c_D$ are some coefficients that depend on how these couplings are generated, and $f_\phi$ is the coupling. Through out the text, we denote $\phi$ as the scanner field. 
If the universe contains non-relativistic baryons and DM, then the finite density potential of $\phi$ is
\bea
V(\phi) = \mB(\phi)\, \nB+ \mDM(\phi)\, \nDM\,. 
\eea
Here we include only the finite-density contributions from baryons and DM in the effective potential.
The scanner’s zero-temperature potential $V_0(\phi)$ is assumed negligible during relaxation (it is generated later by dark-sector confinement, see Section~\ref{sec:scanner_potential}). 
At the minima of the potential $\partial V(\phi)/\partial\phi=0$, and we obtain
\bea
\left. \frac{\rho_{\rm DM}}{\rho_{\rm B}}\right|_{\phi_{\rm min}} \!\!\!\!\simeq  -\frac{c_B}{c_D}\,.
\label{eq:baryon_DM_phi}
\eea
Thus, the ratio of the relic densities is obtained by the dynamics of a field $\phi$ which is set by the ratio of $c_D$ and $c_B$. As long as $c_D$ and $c_B$ are naturally of order the same, then the baryon and DM energy densities are relaxed to similar values.

A key challenge is to realize the required relation between the baryon and dark sector masses in a UV-complete setting. In this work we focus on the QCD axion as a dark matter. Since the mass of the axion is tied to QCD, scanning the axion mass generally scans the masses in baryon sector as well. For the vanilla QCD axion both masses change in the same direction, resulting in no minimum. However, for composite axion scenarios the relation becomes highly predictive. In particular $\Omega_{\rm DM}/\Omega_{\rm B}$ is fixed by a ratio of $\beta$ functions and takes discrete values. Remarkably, in the minimal composite axion model this prediction depends on a single integer parameter $N$, and the choice $N=8$ yields $\Omega_{\rm DM}/\Omega_{\rm B}=5.33$, in excellent agreement with the observed value.  
 
 This framework has several distinctive tests. Improved determination of $\Omega_{\rm DM}/\Omega_{\rm B}$ directly tests the prediction, while progress on lattice determination of the proton mass dependence on the high energy QCD gauge coupling would sharpen the prediction of the model. Additionally the required coupling between the scanner field and baryons motivates more traditional tests such as equivalence principle and fifth-force experiments. 

The paper is organized as follows. 
In Section~\ref{sec:Model}, we present the full field content and master Lagrangian and explain how the $\phi$ coupling to the baryons and DM is generated. 
For our model to work, we need to follow a specific cosmological timeline. 
We chronologically discuss the timeline starting from the end of the inflation till the reheating in Section~\ref{sec:cosmo_timeline}. 
We discuss various constraints on our model in Section~\ref{sec:constarints}, and we conclude  the paper in Section~\ref{sec:conclusion}.

\section{Model} \label{sec:Model}

\begin{table}[t] 
\centering
 \begin{tabular}{|c | l l|} 
 \hline
 Field & $SU(3)\times U(1)_Y \times SU(N) \times U(1)_B$  &  Spin \\ [0.5ex] 
 \hline
 $\phi$ & $(1,0,1,0)$  & $0$ \\ 
 $a$ &  $(1,0,1,0)$  &   $0$ \\
 $\chi$ & $(1,0,N,0)$ &   $1/2$ \\
  $\xi$ & $(1,0,N,0)$ &   $1/2$ \\
   $\psi$ & $(3,0,N,0)$ &   $1/2$ \\
 $\Phi$  &    $(1,0,1,2)$ & $0$ \\
    $S$  &    $(3,2/3,1,-2/3)$ & $0$ \\
 $\Phi_E$  &    $(1,0,1,0)$ & $0$ \\
 $\psi_R$  & $(1,0,1,-1)$  &  $1/2$ \\ [1ex] 
 \hline
 \end{tabular}
 \caption{Fields, associated representations and spin. SU(2) representation is omitted for brevity.}\label{Table}
\end{table}

\subsection{Overview}
In this section we outline the model which can be organized in the following way
\bea
\mathcal{L} &=& \mathcal{L}_{\rm SM} + \mathcal{L}_{\rm relax} 
+ \mathcal{L}_{\rm SU(N)}+ \mathcal{L}_{\rm AD} \\
&+& \mathcal{L}_{\rm scan} + \mathcal{L}_{\Phi_E}\,, \nonumber
\eea
where each sector contains 
\bea\label{eq: sectors}
\mathcal{L}_{\rm relax} &\supset& \mB(\phi)  \bar{B} B + \frac{1}{2} m^2_{a}(\phi)a^2\,, \nonumber\\
\mathcal{L}_{\rm SU(N)} &\supset& \bar\chi \left( i\slashed{D} + M_\chi+y_\chi\phi\right)  \chi\  + \bar\psi i\slashed{D}\psi \nonumber + \bar\xi i\slashed{D}\xi  \\
\mathcal{L}_{\rm AD} &\supset& \vert \partial \Phi \vert^2  - m_\Phi^2 (\Phi^\dagger\Phi) - \epsilon\,m_\Phi^2 (\Phi^2+\Phi^{\dagger 2}) \nonumber\\
&& + y_1 \Phi \psi_R \psi_R+ y_2 \psi_R u_R\, S^\dagger  \\
&&+ \epsilon_{pqr}y_3 S^{p}\, d^{q}_R d^{r}_R \,, \nonumber\\
\mathcal{L}_{\rm scan} &\supset& \frac{g_D^2}{32\pi^2}\frac{\phi}{F}G_D\tilde G_D + \frac{r_X}{4}\frac{\phi}{F}X_{\mu\nu}\tilde X^{\mu\nu}\,, \nonumber\\
\mathcal{L}_{\Phi_E} &\supset& \kappa_E |H|^2 \Phi_E - \frac{1}{2} M_{\Phi_E}^2 \Phi_E^2 \,. \nonumber
\eea

For brevity we list only the interaction terms relevant for cosmology. Each sector has the following role: 
\begin{itemize}
    \item $\mathcal{L}_{\rm SM}$ includes the Standard Model contributions, 
    \item  $\mathcal{L}_{\rm relax}$ includes finite density potential responsible for setting $\rho_{\rm DM}/\rho_{\rm B}$, see Section \ref{sec:phi_dynamics}
    \item $\mathcal{L}_{\rm SU(N)}$ generates the composite axion and relates $f_a$ and $\Lambda_{\rm QCD}$ to $\Lambda_N$, see Section \ref{sec:phi_scales}
    \item $\mathcal{L}_{\rm AD}$ gives rise to early baryon asymmetry with low SM temperature, see Section \ref{sec:Baryogenesis}
    \item  $\mathcal{L}_{\rm scan}$ provides late-time periodic potential for the scanner field that freezes value of $\phi$, see Section \ref{sec:scanner_potential}
    \item $\mathcal{L}_{\Phi_E}$ controls the Hubble expansion and reheats the SM before BBN, see Section \ref{sec:end of inflation}.
\end{itemize}
 
Details of every sector are provided in Sections \ref{sec:phi_scales} and \ref{sec:cosmo_timeline}.

The model is specified by the discrete choice of $N$ (and $D$ for the scanner sector), the renormalizable couplings and masses appearing in Eq. \eqref{eq: sectors}, and cosmological initial conditions (e.g. reheaton/scanner branching ratios).
A convenient independent parameter set is:
\begin{itemize}
\item \textbf{Gauge/field-content:} $N$ (and $D$), field content in Table~\ref{Table}.
\item \textbf{Composite-axion sector:} $M_\chi,\,y_\chi$.
\item \textbf{AD sector:} $m_\Phi,\,\epsilon,\,y_1,\,y_2,\,y_3$.
\item \textbf{Reheating:} $M_{\Phi_E},\,\kappa_E$.
\item \textbf{Scanner sector:} $D,\,g_D,\,F,\,r_X,\,\Lambda_\phi$, and $\alpha_X,\,M_D$ for a concrete glueball-decay completion.
\end{itemize}

For convenience we summarize the main scales appearing in the construction in Table \ref{tab:scales_summary}. 

 \begin{table}[t]
\centering
\begin{tabular}{|c|p{3.4cm}|p{3.3cm}|}
\hline
Scale & Role & Typical value  \\
\hline
$\Lambda_N$ & Confinement scale of the $SU(N)$ sector that gives the axion decay constant $f_a$ (Eq.~\eqref{eq:fa_less_fphi}) & Much above $\Lambda_{\rm QCD}$, similar as $f_a$  \\
\hline
$f_a$ & QCD axion decay constant & final value $f_a \sim 10^8$--$3\times10^{11}\,\GeV$ \\
\hline
$\Lambda_{\rm QCD}$ & QCD confinement scale & Final value $\sim 100\,\MeV$ \\
\hline

$f_\phi$ & Scanner coupling scale controlling the $\phi$-dependence of masses &  $f_\phi/c_B\sim 10^{13}\,\GeV$ \\
\hline
$F$ & Effective scale of the late-time periodic scanner potential & $F\sim 10$--$30\,\TeV$ \\
\hline
$\Lambda_\phi$ & Confinement scale of the scanner sector & a few $\GeV$ \\
\hline
\end{tabular}
\caption{Summary of the main scales appearing in the model.}
\label{tab:scales_summary}
\end{table}

\subsection{$\phi$ Dependence of Scales}\label{sec:phi_scales}

In this section, we construct a UV complete model that generates the $\phi$ coupling to baryons and DM as written in Eq.~\eqref{Eq:baryon_DM_phi}. 
We show that the ratio of $c_B/c_N$, which sets the ratio of relic baryon to DM energy densities, can be obtained naturally in the models of composite axions. 
Moreover, the ratio is fixed, and dictated by the gauge charges of the confined fermions. Finally, our UV model will eventually need to be modified as it does not suppress the vacuum potential of $\phi$ and simply assumes that it has been fine tuned to zero\footnote{For instance, a $Z_N$ symmetry can be employed to naturally suppress contributions from such a potential, as demonstrated in Refs. ~\cite{Hook:2018jle,Brzeminski:2020uhm}}.

The composite axion model we consider is the very first composite axion model considered in Ref.~\cite{Kim:1984pt}, see Appendix~\ref{app:composite_axion} for a detailed discussion.
The confining gauge group that gives the axion is $G = SU(N)$ with massless vector-like
fermions $\psi$ and $\xi$ transforming under $SU(N)$ and $SU(3)$ as $\psi = (N,3)$ and $\xi = (N,1)$. 
$SU(N)$ confines at a scale $\Lambda_N$ much higher than $\Lambda_{\rm QCD}$ and gives the axion in the form of the color singlet combination of $a\sim (\bar\psi\gamma^5\psi-3\bar\xi\gamma^5\xi)$.  
Due to RG running, the 
the IR value of the QCD coupling at 1-loop depends on $\Lambda_N$ as
\bea
\Delta \left(\frac{2\pi}{\alpha_s}\right) = - \frac{2N}{3} \log\left(\frac{\Lambda_N}{\Lambda_{\rm UV}}\right)\,.
\eea
Thus the QCD scale obtains $\phi$ dependence through $ \Lambda_N$ as
\bea
\!\!\!\!\!\!\!\!\!\!\!\!\!\!\!
\Lambda_{\rm QCD}(\phi) \! \propto \!  \left(\Lambda_N(\phi)\right)^{2N/(3\beta_3)} \equiv \Lambda_{\rm QCD}(0) e^{c_B\phi/\fphi},
\label{eq:mB_phi}
\eea
where $\beta_3$ is the IR beta function of $SU(3)$ and we have parameterized the $\phi$ dependence of $\Lambda_N$ as $\Lambda_N(\phi)=\Lambda_N(0)\exp(c_N\phi/f_\phi)\,$.  We obtain 
\bea
r=\frac{c_N}{c_B} = \frac{27}{2N}\,,
\label{eq:r_expression}
\eea
for $\beta_3=9$. 
Consequently, the $\phi$ dependence of $f_a$ is given by 
\bea \label{eq: fa phi}
f_a(\phi) = f_a(0)\exp{\left(c_N\phi/f_\phi\right)}\,,
\eea
then $c_N$ parameterizes how the scanner changes the axion decay constant.

Since $m_a\propto \Lambda_{\rm QCD}^{3/2}/f_a$, Eqs. \eqref{eq:mB_phi} and \eqref{eq: fa phi} imply $c_D=3c_B/2-c_N$, which combined with Eq.\eqref{eq:baryon_DM_phi} then gives 
\bea
\!\!\!\!\!\!\!\!
\frac{\rho_{\rm DM}}{\rho_{\rm B}} = \frac{2}{2r-3} = \frac{2N}{27-3N}\,,
\label{eq:robs_N}
\eea
even without determining $c_B$ and $c_N$ independently.

\begin{figure}
    \centering
    \includegraphics[width=\columnwidth]{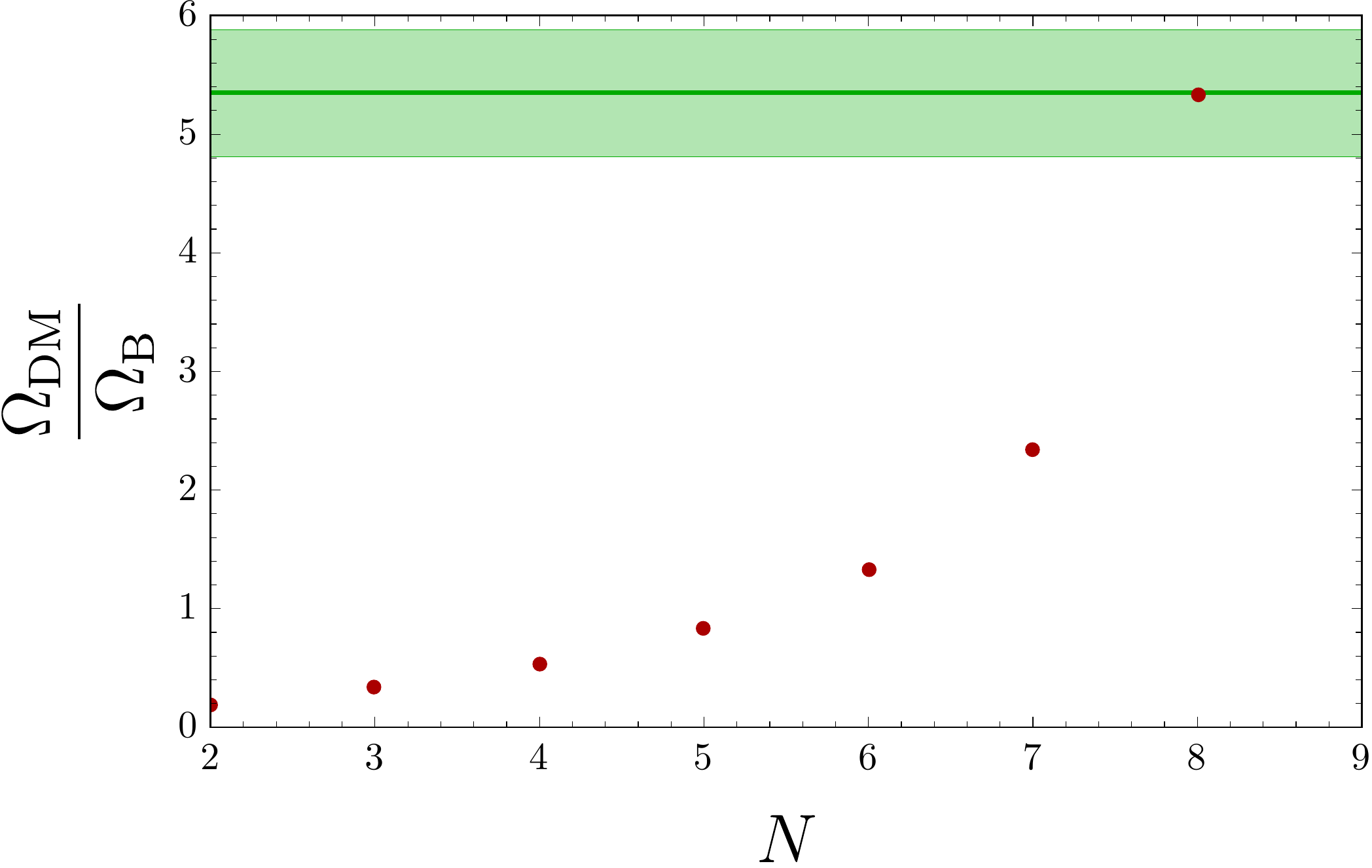}
    \caption{Plot of $\Omega_{\rm DM}/\Omega_{\rm B}$ as a function of $N$. 
    The red dots represent the theoretical predictions obtained using Eq.~\eqref{eq:robs_N}. 
    The shaded green band shows the measured value of 
    $5.36$ (depicted by the green line)~\cite{Planck:2018vyg} with $10\%$ error bar. 
    }
    \label{fig:robs_N}
\end{figure}

To generalize the expression for $r$, notice that all the fermions that are charged under both $SU(3)$ and the confined gauge group affect the IR value of $\alpha_s$. 
For $n_{\psi}$ number of vector like fermions in the ($R_{1,\psi}$,  $R_{2,\psi}$) representations under $SU(3)$ and $G$ respectively, we obtain  $\Lambda_{\rm QCD}(\phi)\propto (\Lambda_N(\phi))^{Q_N/\beta_3}$, and $r=\beta_3/Q_N$ with   
\bea
Q_N = \frac{4}{3}\sum_{\psi} n_\psi C(R_{1,\psi}) d(R_{2,\psi})\,,
\eea
where, $d(R)$ and $C(R)$ denote the dimension and the normalization of the representation $R$ respectively~\cite{Peskin:1995ev}. 
As discussed previously, for the scanner to find minimum, we require the $c_D$ and $c_B$ to have opposite sign. 
Thus, we require $r> 3/2$, and this translates to,
\bea
\frac{c_N}{c_B}>\frac{3}{2}\Rightarrow Q_N < 6\,. 
\eea   
At the same time, subject to the caveats mentioned before, the observed value of $\rho_{\rm DM}/\rho_{\rm B}$ should be reproduced to within $10\%$. 
Thus, $Q_N$ needs to be $\sim 5.3$ with $\sim2\%$ accuracy. 
To illustrate the point, let us consider the model outlined before. 
In this case, $Q_N = 2N/3$ as $G=SU(N)$, and $n_\psi=1$. 
In Fig.~\ref{fig:robs_N}, we plot $\Omega_{\rm DM}/\Omega_{\rm B}$ as a function of $N$. 
The red dots represent the theoretical predictions obtained using Eq.~\eqref{eq:robs_N}. 
The shaded green band shows the measured value of the ratio of cold DM and baryon energy density of $5.36$~\cite{Planck:2018vyg} with $10\%$ error bar. 
For $N=8$, the theoretical prediction falls within the uncertainty of the measurement. 
Thus, we set $N=8$ for the rest of the paper.\footnote{
The requirement to predict $\rho_{\rm DM}/\rho_{\rm B}$ within 10\% of the measured value is $Q_N=16/3$. 
This can be achieved by having two fermions transforming under the fundamental representation under both $SU(3)$ and $G=SU(4)$ and/or one fermion which is in the adjoint representation of $G=SU(3)$ and fundamental representation of $SU(3)$ in addition to the discussed case of one fermion under the fundamental representation of both $G=SU(8)$ and $SU(3)$.  
Any choice with $Q_N=16/3$ keeps the analysis unchanged.}

\subsection*{$\phi$ dependence of the Dark Confinement scale }

As argued above, how the $SU(N)$ confinement scale obtains $\phi$ dependence is not important, it is only the ratio of $c_B/c_N$ that is fixed by the previous model. 
For completeness, here we sketch a simple model to obtain the $\phi$ dependence of $\Lambda_N$.

Let us consider a vector-like fermion $\chi$ with mass $M_\chi$ in the $T_R$ representation of $SU(N)$ that interacts with $\phi$ as
\bea
\mathcal{L}\supset  \left( M_\chi+y_\chi\phi\right) \bar\chi \,\chi\,,
\eea 
with interaction strength $y_\chi$. 
For $M_\chi\gtrsim \Lambda_N$, we integrate out $\chi$ to generate $\phi$ dependence of the $SU(N)$ gauge coupling. 
At one loop order, we obtain the $\phi$ dependence of $\Lambda_N$ for $n_\chi$ number of fermions as,
\bea
\!\!\!\!\!\!\!\!\!\!\!\!\!
\Lambda_N \propto \left( M_\chi+y_\chi\phi \right)^{\!\!\frac{(4 n_\chi T_R/3)}{\beta_N}}  \approx \Lambda_N(0) e^{c_N\phi/f_\phi}\,,
\eea
where in the final step we assume small field limit $y_\chi\phi\ll M_\chi$ and use $\exp(x) \equiv \lim_{N\to \infty}(1+x/N)^N$ to obtain approximate exponential form with $f_\phi\equiv M_\chi/y_\chi$. 
The exponential form simplifies calculations but is not necessary for the mechanism to work.
We defined $\beta_N$ to be the low energy $\beta$-function of $SU(N)$, 
\bea
c_N =\frac{4 n_\chi T_R/3}{\beta_N}\,,\,\,{\rm and}\,\, f_\phi = \frac{M_\chi}{y_\chi}\,.
\eea
In our case, the minimal set-up has 4 light fermions in the fundamental representation of $SU(N)$. 
Thus, we get $\beta_8 = 80/3$, and $c_8 = n_\chi T_R/20$. 
At the same time, as $f_\phi\gtrsim M_\chi\gtrsim \Lambda_N$, the axion decay constant is constrained to be  
\bea
\Lambda_N = \frac{N}{\sqrt{6}} f_a \lesssim f_\phi\,.
\label{eq:fa_less_fphi}
\eea
As long as the $\phi$ dependence of the scales are generated by integrating out heavy states, the above requirement is generic, and independent of the gauge groups, and representations.

\section{Cosmological Timeline}\label{sec:cosmo_timeline}

\begin{figure}
    \centering
    \includegraphics[width=\columnwidth]{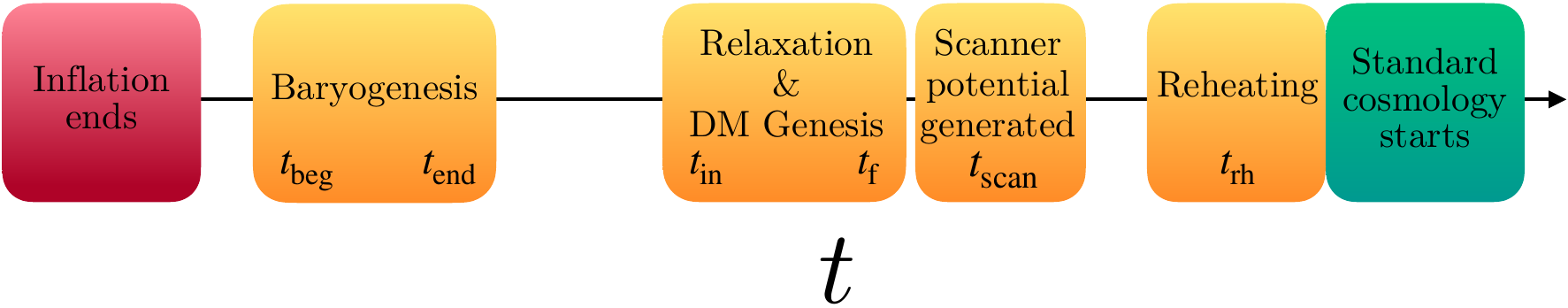}
    \caption{Schematic diagram of the cosmological timeline. 
    Each blob represents a major phase in time: (left to right) end of inflation, baryogenesis, relaxation, phase transition of the scanner sector, and reheating. After reheating, the standard cosmology proceeds. 
    The time stamps inside the blobs represent the beginning (left) and end (right) of each cosmological phase.   
    A light reheaton drives the expansion of the universe at the end of inflation till reheating. This era is marked by the yellow colored blobs. 
    }
    \label{fig:cosmo_timeline}
\end{figure}

In this section we describe cosmological set up that we are considering. 
The cosmological history is illustrated in Fig.~\ref{fig:cosmo_timeline} and consists of several major phases; the end of inflation, the creation of the baryon asymmetry, the relaxation phase, the creation of dark matter, the generation of the scanner potential, and reheating. 

Very briefly, our cosmological history starts with the end of inflation where the inflaton decays into a light reheaton as well as having a small branching ratio into a dark ``scanner" sector. The reheaton is the dominant energy density of the universe and eventually reheats the SM. The scanner sector will be responsible for the mass of $\phi$.

After the end of inflation, comes the formation of the baryon asymmetry and eventually dark matter.  We will take Affleck-Dine baryogenesis and use the misalignment mechanism to produce dark matter.

Afterwards, relaxation occurs.  $\rho_{\rm B}/\rho_{\rm DM}$ is dynamically adjusted to its observed value.
The last step of relaxation is when the scanner sector confines and gives a potential for $\phi$.  This potential fixes in place the value of $\rho_{\rm B}/\rho_{\rm DM}$.

Finally, the reheaton decays and the SM is reheated and standard cosmology occurs.
In what follows, we give a more detailed description of these epochs in chronological order.

\subsection{End of Inflation}\label{sec:end of inflation}

After the end of inflation, the inflaton mostly decays into reheaton fields $\Phi_E$ whose self interactions are small enough that they do not thermalize, and whose mass is small enough that they are produced relativistically.  The bath of $\Phi_E$ particles drives the expansion of the Universe. Eventually while still relativistic, $\Phi_E$ will decay into Standard Model particles and reheat the universe.

The requirement that something other than the Standard Model (SM) dominates the energy density of the universe comes from two considerations. 
First, it is difficult to obtain non-relativistic protons when the SM is hot. 
Second, as discussed in more detail at the end of Sec.~\ref{sec:QCD_axion_DM}, when the SM plasma is hot, the thermal loop induced $\phi$ potential can easily be more dominant than the small non-relativistic baryonic contribution. 
In that case, either the scanner does not successfully scan the energy densities, or the thermal contributions draw $\phi$ to a minima where $\rho_{\rm B}/\rho_{\rm DM} \approx 0$; thus spoiling the relaxation.

To successfully obtain BBN and a post BBN universe, we require $\Phi_E$ to dump its entropy to the SM before few $\MeV$ temperature. 
We consider a $\Phi_E$ interaction with the SM Higgs $|H|$ as 
\bea
\mathcal{L}\supset \kappa_E |H|^2 \Phi_E\,,
\eea
where $\kappa_E$ is the interaction strength. 
With a boost factor of $\gamma_{E}=10^3$ at decay, and $\kappa_E\simeq 10^{-7}\GeV$, we obtain a decay width of
\bea
\Gamma_{\Phi_E} \simeq \frac{\kappa_E^2}{8\pi M_{\Phi_E}\gamma_E}\simeq H(\Trh \simeq 8.4\MeV)\,,
\eea
for $M_{\Phi_E}=10\TeV$. $H(T)$ is the Hubble scale at temperature $T$. 
Thus $\Phi_E$ decays to the SM, and the universe is reheated at $\Trh \simeq 8.4 \MeV$, where we have defined the reheat temperature to be the SM temperature at which $\rho_{\Phi_E} = \rho_{\rm SM}$.

\subsection{Generating Baryon Asymmetry --  Affleck Dine Mechanism}\label{sec:Baryogenesis}

Sometime after the inflaton decays to the reheaton, baryogenesis happens. 
We consider the production of the baryon asymmetry via the Afleck-Dine (AD) mechanism~\cite{Affleck:1984fy}.

We consider Affleck--Dine baryogenesis because, in the cosmological history relevant for our mechanism, one needs not only a baryon asymmetry but also a sufficiently large baryon density before the late entropy injection from reheating. This helps ensure that during relaxation the baryon-induced finite-density potential can dominate over SM thermal contributions, in particular the pion-induced contribution. Other baryogenesis mechanisms may also be viable, but typically would require additional model building and/or a modified cosmological history to achieve the same parametric regime.

Consider a complex scalar $\Phi$ that carries a global baryon number charge of $Q_\Phi$ with the following potential 
\bea
V(\Phi,\Phi^\dagger)= m_\Phi^2 (\Phi^\dagger\Phi) + \epsilon\,m_\Phi^2 (\Phi^2+\Phi^{\dagger 2})\,,
\eea
where $m_\Phi$ is the mass of $\Phi$, and a small real $\epsilon$ softly breaks the symmetry~\footnote{ 
A symmetry preserving quartic term can also be considered in the potential. 
However for small field value i.e $\Phi\lesssim m_\Phi/\sqrt{\lambda_\Phi}$ where $\lambda_\Phi$ being the quartic coupling, the mass terms dominate, and the quartic interaction can be neglected. 
See~\cite{Lloyd-Stubbs:2020sed,Mohapatra:2021aig} for an elaborate discussion.}. 
The resulting asymmetry is generated in the standard AD way from the dynamics of the misaligned condensate and is maximized when the decay rate is comparable to the oscillation splitting. Since only the final asymmetry estimate is needed for the cosmological constraints below, we defer the derivation to Appendix~\ref{app:AD_details}.

The total maximum co-moving baryon asymmetry $\eta_s$, is given by 
\bea
\!\!\!\!\!\!\!\!\!\!\!\!\!\!
\eta_s\equiv\left.\frac{n_{\rm tot}}{s}\right|_{{\tend}}
\!\!\!\! \simeq \frac{Q_{\Phi}}{2}
\frac{ n_{\rm AD}(\tbeg)}{s(t_{\rm ref})}
\left(\frac{\abeg}{R_{\rm ref}}\right)^3 
\,,
\label{eq:eta_S}
\eea
where, $s(t_{\rm ref})=5.89\times 10^{-7} \GeV^3$ is the entropy density at a reference time $t_{\rm ref}$ after reheating has completed, with a reference temperature of $T_{\rm ref}=5\MeV$ and a corresponding scale factor of $R_{\rm ref}/\arh = 2$.  
Using $3H(\abeg)=m_{\Phi}$, we finally obtain
\bea
\!\!\!\!\!\!\!\!\!
\eta_s
\simeq 10^{-10}
\left(\frac{ \rho_{\rm AD}(\tbeg)}{10^{30}\GeV^4}\right)\left(\frac{ 100 \TeV}{m_\Phi}\right)^{5/2},
\label{eq:etas_final}
\eea
where $\rho_{\rm AD}(\tbeg)=m_\phi n_{\rm AD}(\tbeg)$ is the initial energy density of the AD field, and we take $Q_\Phi=2$. 
For the benchmark AD dynamics summarized in Appendix~\ref{app:AD_details}, the onset time of baryogenesis is
\bea
\frac{\tbeg}{\trh} = 1.4\times 10^{-27} \left(\frac{100\TeV}{m_\Phi}\right).
\eea 
As the total maximum co-moving asymmetry is determined by the quantities at $\tbeg$, an observational constraint on $\eta_s\lesssim 10^{-10}$~\cite{ParticleDataGroup:2020ssz} can be translated to an upper bound on the energy density of the AD field as (using Eq.~\eqref{eq:etas_final}),
\bea
\rho_{\rm AD}(\tbeg)\lesssim  10^{30} \GeV^4 \left(\frac{m_\Phi}{100\TeV}\right)^{5/2}. 
\eea

Baryogenesis ends when the decay rate of the AD field becomes comparable to the Hubble scale i.e.  $H(\tend)\equiv\Gamma_\Phi$. 
To calculate the decay width, we consider the following interaction 
\bea
\!\!\!\!\!\!\!\!\!\!\!\!\!\!
\mathcal{L} \supset y_1 \Phi \psi_R \psi_R+ y_2 \psi_R u_R\, S^\dagger + \epsilon_{pqr}y_3 S^{p}\, d^{q}_R d^{r}_R \,,  
\eea
where, $p$, $q$, $r$ are color indices, $y_i$s are the yukawa couplings, $\psi_R$ is a right-handed color singlet fermion with $U(1)_B$ charge of $-1$, $d_R$ ($u_R$) denotes the SM right-handed down (up) type quarks, and $S$ is a scalar which transforms under the fundamental representation of $SU(3)$ with $U(1)_B$ charge of $-2/3$ and hypercharge of $2/3$. 
For notational simplicity we omit the flavor indices on the quarks and possibly on the yukawa couplings as well. 
$\Phi$'s decays cascade into $uddudd$ (2 neutrons), and the above interactions fixes the $U(1)_B$ charge of $\Phi$ to be $2$. 
To simplify the calculation, we take the mass hierarchy to be  $m_\Phi/2\gtrsim m_\psi\gtrsim m_S$. 
For the $\Phi\to \psi_R\psi_R $ decay, the width can be written as, 
\bea
\!\!\!\!\!\!\!\!\!
\Gamma_{\Phi} = \frac{y_1^2\,m_\Phi}{8\pi} \left(1-\frac{4 m_\psi^2}{m_\Phi^2}\right)^{1/2}\left(1-\frac{2 m_\psi^2}{m_\Phi^2}\right)\,.
\eea
For $\psi\to u S$, and $S\to dd$ decay, we consider the SM quarks to be mass-less and obtain
\bea
\!\!\!\!\!\!\!\!\!\!\!\!\!\!\!\!\!\!
\Gamma_\psi = \frac{3 y_2^2\,m_\Psi}{16\pi} \left(1-\frac{m_S^2}{m_\psi^2}\right)^{2}\!\!\!\,{\rm and}\,\,\, \Gamma_S = \frac{y_3^2\,m_S}{8\pi}\,, 
\eea
respectively. 
Note that, for $\Gamma_S$, we consider $S$ decaying into quarks of two different flavors. 
If $\Phi\to\psi\psi$ is the rate determining step in the cascade decay of $\Phi$, we obtain the duration of baryogenesis as
$\tbeg/\tend= 3\gamma$ by using $H(\tend)=\Gamma_\Phi$. 
A hierarchical decay widths of $\Gamma_\Phi\ll \Gamma_\psi\ll \Gamma_S$, can be obtained by choosing appropriate masses and couplings of the fields as discussed in the following. 
\\

Finally, we provide a benchmark point for the our baryogenesis model. 
Note that the exact values of the parameters are not important, as long as the hierarchy is maintained. 
We choose $m_\Phi=100\TeV$, $m_\psi = 45\TeV$, $m_S=40\TeV$, $y_1=3.7\times 10^{-4}$, and $y_2 =y_3=0.1$. 
With these set of parameters, we obtain $\Gamma_\Phi \simeq 0.3 \MeV$, $\Gamma_\psi \simeq 1.2 \GeV$, and $\Gamma_S\simeq 16 \GeV$. 
Thus, $\gamma\simeq 0.33 \times 10^{-8}\sim \epsilon$, and the period of baryogenesis lasts $\tbeg/\tend=10^{-8}$. 
\\

For the benchmark above, washout is negligible. The leading baryon-violating processes induced by the $\epsilon$ spurion are suppressed because the SM temperature at the end of baryogenesis is below the heavy-state masses, and electroweak sphalerons modify the estimate by at most an $\mathcal{O}(1)$ factor. The new colored states are also safely above current direct-search bounds for the benchmark choice. Flavor constraints depend on the Yukawa texture and can be avoided with appropriate flavor assignments, so they do not affect the cosmological discussion below.

\subsection{Relaxation Phase}\label{sec:phi_dynamics}
\begin{figure}
    \centering
    \includegraphics[width=\columnwidth]{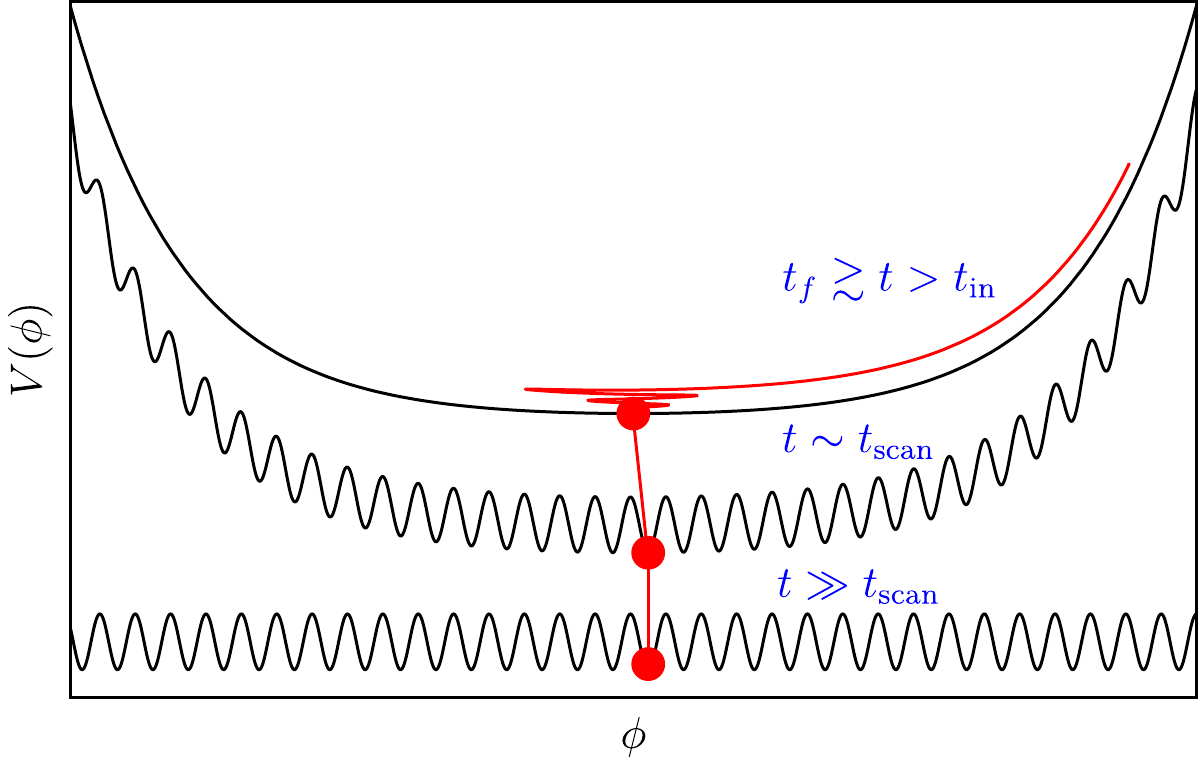}
    \caption{An example showing the evolution of the scanner field and its potential, assuming $t_{\rm osc} < t_{\rm in}$. At $t=t_{\rm in}$ the scanner starts to roll and reaches the minimum at $t=t_f$. Later, at $t=t_{\rm scan}$, the temperature dependent potential turns on and locks in the vacuum expectation value of the scanner and the ratio of DM and baryon abundances.
    }
    \label{fig:field_evolution}
\end{figure}

After baryogenesis ends, relaxation can start once the baryons are non-relativistic. 
In this section, we briefly discuss the dynamics of the scanner field, $\phi$, which relaxes the baryon and DM energy densities as outlined in~\cite{Brzeminski:2023wza}. 
A schematic representation of the scanner evolution is shown in Fig.~\ref{fig:field_evolution}.  
\\

In Section~\ref{sec:phi_scales}, we constructed a model where both the QCD scale $\Lambda_{\rm QCD}$ and the axion decay constant $f_a$ depend on $\phi$. For QCD axion DM this implies a $\phi$-dependent axion mass through $m_a\propto \Lambda_{\rm QCD}^{3/2}/f_a$. 
Following Eq.~\eqref{eq:mB_phi}, we write the interaction of $\phi$ with the baryons and DM as,   
\bea
\mathcal{L}\supset \mB(\phi)  \bar{B} B + \frac{1}{2} m^2_{a}(\phi)a^2\,,
\eea
where, $B$ denotes a nucleon (baryon), and $\mB(\phi)\simeq 10\,\Lambda_{\rm QCD}(\phi)$\footnote{As a normalization, we are taking the final values of $\Lambda_{\rm QCD}=100 \MeV$, and the baryon mass $m_{\rm B}=1\GeV$.}. 
We simplify the QCD axion potential, and write the mass term as $m_{a}=m_{a}(0)\exp(c_D\phi/f_\phi)$ with $c_D= 3c_B/2-c_N = c_B (3/2-r)$. 
At the beginning of the relaxation the baryons and possibly DM~\footnote{The DM is produced only after the axions start to oscillate. 
We require the oscillation starts before the end of relaxation.} are non-relativistic and we have 
\bea
V(\phi)= \mB(\phi) n_B+ m_{a}(\phi)\na\,. 
\eea
The equation of motion (EOM) of $\phi$ can be written as,
\bea
\ddot\phi &+& 3H\dot\phi +  \frac{c_B}{\fphi}\mB(0)e^{c_B\phi/\fphi}n_B\nn\\
&+&\frac{c_D}{\fphi}m_a(0)e^{c_D\phi/\fphi}\na =0\,.
\label{eq:phi_EOM}
\eea 
To solve the above EOM, we define $\varphi(t) = e^{c_B\phi/\fphi}$. 
In the radiation dominated universe, we find an attractor solution in the form of 
\bea
\varphi(t) = \varphi(t_i)\left(\frac{t_i}{t}\right)^{1/2}\propto 1/R(t)\,,
\eea
for $c_B\rho_{\rm B}(t)\gtrsim c_D\rho_{\rm DM}(t)$ with the field value at some reference time $t_i$ satisfying  
\bea
\frac{c^2_B}{\fphi^2}\mB(0)n_B(t_i) \varphi(t_i)  = \frac{1}{4t_i^2}\,.
\eea
In other words, as long as the initial conditions are such that $\phi$ is frozen out, $m_\phi \lesssim H$, then it will approach the attractor solution.
Afterwards the $\phi$ field relaxes the baryon mass while keeping a mass 
\bea
m_\phi^2(t) = \frac{c^2_B}{\fphi^2}\mB(0)\varphi(t)n_B(t) =H^2(t)\,.
\label{eq:mphi_hti}
\eea 
Note that this scaling solution is valid as long as the baryon energy density drives the $\phi$ potential.
The time (scale factor) when the relaxation starts $\tin$ ($\ain$), can be obtained by solving $m_\phi(\tin)=H(\tin)$. 
\\
 
The scanning of the baryon mass is governed by $\varphi(t)$ during the relaxation phase. As $\varphi(t)\propto 1/R(t)$, $\mB$ is decreasing during the relaxation as, 
\bea
\!\!\!\!\!\!\!\!\!\!\!\!\!\!
\mB(\phi)=\mB(0)e^{c_B\phi/\fphi} = \mB(0)\varphi(t)\propto 1/R(t). 
\label{eq:mB_scan}
\eea
The field velocity during the scanning phase is given by $\dot\phi(t)/(\fphi/c_B)=H(t)$. During this time, the axion mass is being scanned mildly and is given by   
\bea
m_a(\phi)=m_a(0)e^{c_D\phi/\fphi} \propto R(t)^{-c_D/c_B}\,.
\eea 

At some time, $t=t_f$, the DM contribution to the $\phi$ EOM (c.f. Eq.~\eqref{eq:phi_EOM}) becomes important, and the scaling solution breaks down. 
At this point, both dark matter and baryons are important and the potential is minimized around  
\bea
\frac{\rho_a(t_f)}{\rho_{\rm B}(t_f)} = \left.\frac{m_a\,n_a}{\mB\, n_B}\right|_{t_f}= -\frac{c_B}{c_D}\,,
\eea
where, $t_f$ denotes the time when the relaxation ends. 
Thus, the ratio of the baryon and DM energy densities are set to its present value due to the dynamics of the scanner, and set by the ratios of two model parameters. 
At the end of the relaxation, $\phi$ starts to oscillate around its minima. 
The mass of $\phi$ is given by the matter energy density and red-shifts as $m_\phi^2\propto \rho_{\rm B}\propto R^{-3}$. 
The amplitude of such oscillation decreases with $R$ as, 
$\phi\propto m_\phi^{-1/2}R^{-3/2}\propto R^{-3/4}$.
Thus the energy density of $\phi$ quickly red-shifts away as $\rho_\phi\propto m^2\phi^2\propto R^{-9/2}$.

Now we want to comment on the duration of the relaxation phase. 
As $m_{\rm B}\propto 1/R(t)\propto t^{-1/2}$ and the final baryon mass is fixed to $\GeV$, the duration of the relaxation phase is determined by the initial baryon mass as
\bea
\frac{\tf}{\tin} = \left(\frac{\mB(\tin)}{\GeV}\right)^2\,. 
\eea

Next, we estimate what is the typical value of $\fphi$ that we consider. 
Using Eq.~\eqref{eq:mphi_hti} and redshifting backwards from reheating, we obtain $\fphi$ as 
\bea
\!\!\!\!
f_\phi/
c_B =  10^{13}\GeV\left(\frac{\tf}{2.5\times 10^{-7}\,\trh}\right)^{1/4}\!\!. 
\label{eq:fphi_af}
\eea

Relaxation starts in the confined phase of QCD. 
For an initial QCD scale $\mB(\tin)\gtrsim 175 \GeV$, the top quarks confines. 
This changes the Higgs potential and shifts the Higgs vev ($\vew$). 
To simplify matters, we require $\delta\vew/\vew\lesssim \mathcal{O}(4\pi)$, and  obtain a constraint on the initial baryon mass  
as,
\bea
\!\!\!\!\!\!\!\!\!\!\!\!
\mB(\tin) \lesssim 10\,\vew \left(\frac{16\pi\lambda_H}{4\pi y_t}\right)^{1/3}\sim 1.5 \TeV\,,
\eea 
where we use $m_{\rm B}= 10\,\Lambda_{\rm QCD}$, and $\lambda_H=0.13$ and $y_t=0.99$ are the Higgs quartic and the top quark Yukawa couplings respectively.\footnote{
In a generic confined theory, the SM gauge bosons obtain a mass proportional to the pion decay constant~\cite{Pich:1995bw}. 
By requiring the 1 loop contribution to the Higgs mass from the gauge bosons to be small i.e. $\delta m^2_h/m^2_h\lesssim \mathcal{O}(1)$, we obtain a constraint on the initial value of the pion decay constant as, 
\bea
\!\!\!\! 
f_\pi(\tin)\lesssim 4\pi \vew \left(\frac{16\lambda_H}{9g^4+3g'^4}\right)^{1/2}\sim 2.5\TeV,
\eea
where, $g=0.65$, and $g'=0.35$ are the SM gauge couplings. 
As $m_B\sim 4\pi f_\pi$~\cite{Witten:1980sp}, this provides a weaker constraint on the initial baryon mass than what we have considered before.} 
We choose the minimum value of the initial baryon mass $\mB(\tin)= 5\GeV$. 
Thus, the duration of relaxation phase spans from $25 \lesssim \tf/\tin \lesssim  2.25\times 10^{6}$.
\\

So far we have not discussed how DM is produced which is a crucial part of the relaxation dynamics as
the DM contribution to the $\phi$ potential forces the field to find its minima. 
Thus, for the relaxation to work, it is important to make sure DM is produced by the end of relaxation. 
We discuss DM production from the misalignment mechanism in the next section.

\subsection{Dark Matter Genesis }\label{sec:QCD_axion_DM}

In this section, we discuss the misalignment production of the QCD axion. 
As discussed previously, in our case, both the mass and the decay constant of the axion are being scanned during the relaxation period. 
This leads to interesting scaling behavior of the initial misalignment angle.

Before discussing the evolution of the QCD axion in our case, let us first outline the usual scenario. 
The zero temperature axion mass can be written as, 
\bea
m_a^2(0) = \frac{z}{(1+z)^2}\frac{m_\pi^2 f_\pi^2}{f_a^2}\equiv \frac{\chi(0)}{f_a^2}\,,
\label{eq:axion_T_0}
\eea
where, $m_\pi=140\MeV$, $f_\pi=92.7\MeV$, $z\simeq 0.48$ are the pion mass, pion decay constant, and the ratio of the up and down quark masses respectively~\cite{ParticleDataGroup:2022pth}. 
Also, $\chi$ is the topological susceptibility. 
The zero temperature value of $\chi$ obtained using the above formula $\chi(0)= 3.7\times 10^{-5}\GeV^4$ agrees extremely well with its lattice simulated value~\cite{Borsanyi:2016ksw}. 
At finite temperature, the axion mass receives a temperature dependence through $\chi$ as~\cite{GrillidiCortona:2015jxo, Borsanyi:2016ksw}
\bea
\!\!\!\!\!\!\!\! 
m_a^2(T) = \frac{\chi(T)}{f_a^2} \simeq m^2_a(0) \begin{cases} \left(\frac{T_c}{T}\right)^8\,{\rm for}\,\, T\geq  T_c\\
1 \,\,{\rm for}\,\, T\leq  T_c
\end{cases}
\label{eq:axion_T}
\eea
where, $T_c$ is the QCD phase transition temperature.  
\\

In general, to solve the axion EOM with time dependent mass and decay constant, we define 
\bea
\theta(t)\equiv  \frac{a(t)}{f_a(t)}\,.
\eea
In terms of $\theta(t)$, the EOM takes the form of,
\bea
\ddot\theta+ \left[3\,\frac{\dot R}{R}+ 2\,\frac{\dot f_a}{f_a}\right]\dot\theta+ m_a^2(t)\sin\theta=0\,, 
\eea
 and the energy density, $\rho_{a}$ can be written as,
 \bea
\rho_a(t) = f_a^2(t)\left[\frac{1}{2}\dot\theta^2+ m_a^2(t)(1-\cos\theta)\right].
\label{eq:rho_a_exp}
 \eea
For a slowly varying axion mass i.e. $\dot m_{a}/m_a\ll H$, the phase of $\theta (t)$ varies adiabatically, and we get for $t\gtrsim \tosc$,
\bea
\theta(t>\tosc) = \theta_0(t)\cos\left[\int^t dt'\, m_{a}(t')\right]\,,
\eea
where, $\tosc$ is solved by $3H(\tosc)=m_a(\tosc)$, and  $\theta_0(t)$ scales as,  
\bea
\theta_0(t) \propto \frac{1}{m_{a}^{1/2}}  \frac{1}{f_a}\frac{1}{R^{3/2}}\,.
\eea
Aside from simply solving the WKB approximation, this result can be easily understood as $f^2_a(t)$ appears in a very similar manner as the scale factor and thus the amplitude dependence can be read off from the scale factor dependence.
Plugging in $\theta(t)$ to the expression for the energy density, we obtain 
\bea
\!\!\!\!\!\!\!\!\!\!
\rho_{a}(t)\simeq \frac{1}{2}m^2_a(t) f_a^2(t) \theta^2_0(t)\simeq m_a(t) n_a(t).
\eea
Thus the axion number density $n_a$ red shifts  $n_a \propto m_a f^2_a\theta_0^2\propto R^{-3}$ as expected, despite having a time dependent decay constant. 
\\

In our case, the relaxation starts when QCD is already confined. 
However as the QCD scale is being scanned during the relaxation phase, the topological susceptibility changes with time as,
\begin{equation}
\!\!\!
\chi(t) = \chi(\tin) \left(\frac{\ain}{R}\right)^{3} \,\,\,{\rm for}\,\, \tin \lesssim t\lesssim \tf\,.
\end{equation}
During the relaxation phase, $f_a$ decreases with time as 
$f_a(t) \propto e^{c_N\phi/f_\phi}\propto R^{-r}\,$. 
As the scanning of $f_a$ depends on the IR beta function of $SU(3)$ (c.f. Eq.~\eqref{eq:r_expression}), it is tied tightly to the numbers of quark flavor lighter than a given QCD scale. 
In general $r$ is given by $r = (33-2n_0)/16$ for $n_0$ light quarks. 
Thus, $r$ increases during the relaxation, and most importantly, the scanning rate of $f_a$ increases as we approach the end of relaxation. 
For the QCD scale just below the  top quark mass $m_t$, we get $r=23/16$, which increases by $2/16$ as QCD scale drops below the threshold of a quark mass.  
Eventually below the charm quark mass $m_c$, the value of $r$ becomes $27/16$ and sets $\rho_{\rm B}/\rho_{\rm DM}$ in stone. 
In our case, as we take the range of initial baryon mass from $(5\GeV-1.5\TeV)$, the initial QCD scale ranges from $(500\MeV-150\GeV)$. 
Thus, the final value of the axion decay constant can be obtained as, 
\bea
\!\!\!\!\!\!\!\!\!\!\!\!
\frac{f_a(\tf)}{f_a(\tin)} = \left(\frac{\GeV}{m_{\rm B}(\tin)}\right)^{\frac{23}{16}}
\left(\frac{\GeV^2}{10^2\, m_b m_c}\right)^{\frac{2}{16}}\,,
\label{eq:fa_in_mBin_less_mt}
\eea
for $m_b\lesssim \mB(\tin)/10 \lesssim  m_t$, 
\bea
\!\!\!\!\!\!\!\!\!\!
\frac{f_a(\tf)}{f_a(\tin)} = 
\left(\frac{\GeV}{\mB(\tin)}\right)^{\frac{25}{16}}
\left(\frac{\GeV}{10\, m_c}\right)^{\frac{2}{16}}\,,
\label{eq:fa_in_mBin_less_mb}
\eea
for $m_c\lesssim \mB(\tin)/10\lesssim m_b$, and 
\bea
\!\!\!\!\!\!\!\!\!\!
\frac{f_a(\tf)}{f_a(\tin)} = 
\left(\frac{\GeV}{\mB(\tin)}\right)^{\frac{27}{16}}\,,
\label{eq:fa_in_mBin_less_mc}
\eea
for $\mB(\tin)/10\lesssim m_c$. 
We denote the bottom and the charm quark masses as $m_b$ and $m_c$ respectively. 
The upper bound on the axion decay constant at the beginning of relaxation follows from the EFT requirement $\Lambda_N\lesssim f_\phi$ (Eq.~\eqref{eq:fa_less_fphi}, which implies $f_a\lesssim \mathcal{O}(1)\times f_\phi$ in the composite axion model) together with the relation between $f_\phi$ and $t_f/t_{\rm rh}$ fixed by the end of the relaxation (Eq.~\eqref{eq:fphi_af}). It reads
\bea
\!\!\!\!\!\!\!\!\!\!\!\!\!\!\!
f_a(\tin) \lesssim 3\times 10^{12}\GeV\left(\frac{\tf}{2.5\times 10^{-7}\,\trh}\right)^{1/4}\,.
\label{eq:fa_fphi_final}
\eea
Thus, given how much time there is before the end of relaxation and reheating, the maximum $f_a(\tf)$ is obtained by using Eq.~\eqref{eq:fa_in_mBin_less_mt},~\eqref{eq:fa_in_mBin_less_mb}, or~\eqref{eq:fa_in_mBin_less_mc} in conjunction with the above constraint.    
\\

 Using $m_a\propto \Lambda_{\rm QCD}^{3/2}/f_a$, we see that during relaxation the axion mass scales as $m_a\propto R^{r(t)-3/2}$, where $r(t)\equiv c_N/c_B$ is effectively time-dependent because it is set by the IR QCD $\beta$-function appropriate to the quark content below the instantaneous confinement scale.
Importantly, the prediction for $\rho_{\rm DM}/\rho_{\rm B}$ in Eq.~\eqref{eq:robs_N} refers to the ratio at the end of relaxation, evaluated at the field value where the baryonic contribution self-consistently sits at $m_B\simeq 1\,\GeV$. At that stage the relevant QCD $\beta$-function is the three-flavor one, and $r>3/2$ so the predicted ratio is positive.
The intermediate regimes with $r(t)>3/2$ or $r(t)<3/2$ only control the time dependence of $m_a(t)$ during the scan. In particular, at early times one can have $r(t)<3/2$ so $m_a$ decreases with time, and it changes its behavior once the QCD scale falls below heavy-quark thresholds. 
The axion mass at the end of the relaxation can be obtained as, 
\bea
\frac{m_a(\tf)}{m_a(\tin)} \simeq 2.2\times \left(\frac{\GeV}{\mB(\tin)}\right)^{1/16}\,,  
\label{eq:ma_mB_ls_mt}
\eea
for $m_b\lesssim \mB(\tin)/10\lesssim m_t$, whereas we get
\bea
\frac{m_a(\tf)}{m_a(\tin)} =  1.37\times \left(\frac{\mB(\tin)}{\GeV}\right)^{1/16}\,,
\label{eq:ma_mB_ls_mb}
\eea 
for $m_c\lesssim \mB(\tin)/10\lesssim m_b$. Finally for $\mB(\tin)/10\lesssim m_c$, we obtain 
\bea
\frac{m_a(\tf)}{m_a(\tin)} = \left(\frac{\mB(\tin)}{\GeV}\right)^{3/16}\,. 
\label{eq:ma_mB_ls_mc}
\eea 
Thus the axion mass barely changes during the relaxation. 
Once the relaxation phase ends, neither the QCD scale nor the axion decay constant changes. 
Thus the present day mass of the QCD axion is set by its value at the end of the relaxation i.e. $m_a(0)=m_a(\tf)= \sqrt{\chi(0)}/{f_a(\tf)}=\sqrt{\chi(0)}/{f_a}$. 
After that, the axion energy density redshifts as $\rho_{a} (t\gtrsim \tf)=m_a n_a (t)\propto 1/R^3$, as DM should.  \\

The simplicity of our set up is that, the DM and baryon abundance is fixed at the end of the relaxation. 
Thus, the value of $\theta_0$ at the end of the relaxation can be written as,
\bea
\!\!\!\!\!\!\!\!
\theta_0(\tf)= \sqrt{\frac{2\rho_{a}(\tf)}{\chi(0)}} = \sqrt{\frac{2 r_{\rm obs}}{\chi(0)}}\sqrt{\rho_{\rm B}(\tf)}\,.
\label{eq:theta_0_exp}
\eea
For a given $\tf$, the baryon energy density at the end of the relaxation is
\bea
\!\!\!\!\!\!\!\!\!\!
\rho_{\rm B}(\tf)= 1\GeV \times \eta_s\times s(\Trh)\times  \left(\frac{\arh}{\af}\right)^3\,.
\eea
Combining two above equations, 
we obtain
\bea
\theta_0(\tf) = 1\left(\frac{2.5\times 10^{-7}\,\trh}{\tf}\right)^{3/4}\,, 
\eea
for $\eta_s=10^{-10}$.
By requiring, $\theta_0\lesssim \mathcal{O}(2\pi)$ at all time, we obtain a lower bound on the time when the relaxation needs to end as,  
\bea
\tf/\trh \gtrsim 2  \times 10^{-8}\,.
\label{eq:theta_2Pi}
\eea
Note that,  $\theta_0\propto R^{r/2-3/4}\propto m_a^{1/2}$. 
Thus like $m_a$, at different phases of the relaxation, $\theta_0$ can increase (for $r>3/2$) or decrease (for $r<3/2$) with time. 
However the scanning is negligible as $r$ is very close to $3/2$. 
Thus we find $\theta_0(\tf)\simeq \theta_0(\tin)$ (changing at most by a factor of $\sim 1.2$ for $\mB(\tin)\sim\TeV$). 
However, as the axion is produced at $\tosc\lesssim \tf$ (which we discuss in the next paragraph), the value of $\theta_0$ during the relaxation can be taken as $\theta_0(\tf)$.  
\\

So far we have not commented on the time when the field starts to oscillate $t_{\rm osc}$. 
In our construction, the exact time of oscillation is not important as long as the DM is produced by the end of relaxation i.e. $\tosc\lesssim t_{f}$. 
The time when the axion starts to oscillate is determined by solving  $m_a(\tosc)=3H(\tosc)$. 
Thus the requirement of $\tosc\lesssim t_f$ translates to $m_a(\tosc)\gtrsim 3 H(t_f)$. 
We can rewrite the above inequality as $m_a(t_f)/3 H(t_f)
\gtrsim m_a(t_f)/m_a(\tosc)$. 
During the relaxation, the Hubble scale decreases as $H\propto 1/R^2$ while the axion mass almost remains constant. 
Thus, we take $\tosc\sim \tf$, and obtain a constraint on the axion decay constant as 
\bea \label{Eq:fabound}
f_a(\tf) \lesssim 10^{13} \GeV\left(\frac{\tf}{2.5\times 10^{-7}\,\trh}\right)\,.
\eea
Thus the requirement of axion production during the relaxation leads to a weaker bound (by at least a factor of $10$) than the already considered case of  Eq.~\eqref{eq:fa_fphi_final}. 
Note that, for smaller axion decay constant, $f_a\sim 10^8\GeV$, the axion starts to oscillate even before the beginning of the relaxation. 
In that case the requirement of $\tosc\lesssim t_f$ is trivially satisfied.   
\\

In order for the relaxation to proceed as described in section~\ref{sec:phi_dynamics}, it is critical that $\phi$ does not have any other contributions to its potential.
As $\phi$ couples to the baryons at the tree level, there is a loop induced coupling to the SM photons as well as a coupling to the pions. 
For the SM bath temperature of $T_\gamma$, the photon and electron induced thermal potential of $\phi$ can be written as $V_T(\phi)\sim c_{\gamma}\alpha/(4\pi) \phi\, T_\gamma^4/f_\phi\sim 10^{-6}\,\phi\, T_\gamma^4/f_\phi$ where $c_\gamma$ depends on the relativistic species in the bath~\cite{Quiros:1999jp}.
Meanwhile the pion induced potential for $\phi$ takes the form of $V_\pi(\phi)\sim m^2_\pi(\phi) T_\gamma^2$ when the pions are relativistic. 
For $m_\pi\gtrsim T_\gamma$, despite the Boltzmann suppression, pions can contribute significantly to the $\phi$ potential, and displace it from the special minima
(see Eq~\eqref{eq:pion_NR} and discussion around it)\footnote{As during the relaxation $\rho_{\rm B}\propto 1/R(t)^4$, and the relaxation starts when baryons are non-relativistic, these contributions are important at the end of the relaxation.}.

The final stage of the evolution of $\phi$ is when the scanner potential turns on, preventing thermal corrections from spoiling the relaxation mechanism. We discuss this in the next section.

\subsection{Temperature dependent scanner potential}
\label{sec:scanner_potential}

In this section, we construct a model to generate a temperature dependent scanner potential for $\phi$ which is independent of the ambient baryon and DM density. 
While we could simply write down a vacuum potential for $\phi$, introducing temperature dependence alleviates constraints coming from fragmentation~\cite{Fonseca:2019ypl}.

We consider an axion-like coupling of $\phi$ to a dark gauge group, say $SU(D)$ (denoted as the scanner sector). 
Then similar to the QCD axion, the $SU(D)$ finite temperature instantons would give rise to a temperature dependent potential for $\phi$. 
If confinement in the scanner sector happens after the end of relaxation, $\phi$ quickly settles to the minima closest to where $\rho_{\rm DM} \sim \rho_{\rm B}$. 
The relaxed value of $\rho_{\rm B}/\rho_{\rm DM}$ is not altered as long as there is a (even small) hierarchy between the periodicities of two $\phi$ potentials. 
After confinement, the induced potential for $\phi$ is large enough that SM thermal corrections do not spoil the relaxation mechanism.

To give this sector a non-zero temperature, we initially populate the scanner sector by a small branching ratio coming from the inflaton decay. 
We are assuming that the energy density in the scanner sector is small compare to that of the $\Phi_E$ sector which drives the expansion of the universe. 
To avoid issues of overclosing the universe, a massless dark photon will be introduced allowing for everything to decay.
\\

In this section, we will first discuss the temperature dependent mass of the scanner $\phi$, both how that sector obtains a temperature and how the changing mass affects $\phi$ dynamics.  Next we will discuss how to allow everything to decay so that the universe is not overclosed by any stable particles.

To start, let us consider $\phi$ coupling to $SU(D)$ gauge fields $G_D$ as
\bea
\mathcal{L}\supset \frac{g_D^2}{32\pi^2}\frac{\phi}{F}G_D\tilde G_D\,,
\eea
where $g_D$ is the $SU(D)$ gauge coupling and $\tilde G_D$ is the dual of $G_D$. 
Finite temperature $SU(D)$ instantons generate a potential of the form of 
\bea
V_{\rm scan}(\phi) = \Lambda_{\phi}^4(T_{\rm scan})\cos(\phi/F+\delta)\,,
\label{eq:scanner_potential}
\eea
where, $\Lambda_{\phi}$ depends on the temperature of the scanner sector $T_{\rm scan}$, and $\delta\sim\mathcal{O}(1)$ is a generic phase. 
Without loss of generality, we take the critical temperature of the scanner sector equal to the confinement scale of $SU(D)$ i.e. $T_{\rm scan}^{\rm crit}=T_{\rm scan}(t_{\rm scan})\sim \Lambda_\phi$ where we assume that the scanner sector phase transition happens at time $t=t_{\rm scan}$ with the scale factor $R(t_{\rm scan})=\ascan$. 
We take the form of the temperature dependence as, 
\bea
\Lambda^4_{\rm \phi}(T_{\rm scan}) = \Lambda_\phi^4 \begin{cases}
\left(\frac{R}{\ascan}\right)^\beta\,\,{\rm for}\,\, t\lesssim t_{\rm scan}\\
1\,\,{\rm for}\,\,\, t\gtrsim t_{\rm scan}\,,
\end{cases}
\eea
where, $\beta>0$ captures the thermal dependence of the $SU(D)$ instantons. 
Generalizing the result of $SU(3)$, we obtain $\beta =11D/3 -4$.\footnote{
For $n_f$ number of light flavors under $SU(D)$, we get $\beta =11/3 D-4+n_f/3$. For $n_f=3$ and $D=3$ we obtain the known result of $\beta=8$ for QCD~\cite{GrillidiCortona:2015jxo}.} 
 
Note that, the exact value of the $\beta$ is not important for our discussion, as long as the potential is sufficiently suppressed for $t\lesssim t_{\rm scan}$. 

After the end of relaxation, $\phi$ oscillates around the minimum with a slowly decaying amplitude.
After $t_{\rm scan}$, the value of $\phi$, and hence $\Omega_{\rm DM}/\Omega_{\rm B}$, is locked in place.  For the parameters we choose, the locked in value of $\phi$ is determined by the amplitude of the oscillation at the time.  Thus it is important to know what is the initial value of the oscillation amplitude and how it decays.

By numerically solving for the behavior of $\phi$ after the relaxation period ends, we find that the initial value of the oscillations obeys $\delta (\Omega_{\rm DM}/\Omega_{\rm B}) / (\Omega_{\rm DM}/\Omega_{\rm B}) \approx 20\%$.  Thus if $t_{\rm scan} = t_f$, our prediction for the value of $\Omega_{\rm DM}/\Omega_{\rm B}$ will have error bars of about 20\%.  When $t_{\rm scan} > t_f$, $\phi$ will oscillate for a while before getting locked in place.

As noted previously, after the end of relaxation, the baryon and DM induced energy density of $\phi$ redshifts as $\rho_{\phi}\propto  R^{-9/2}$. 
By defining $\delta\phi=\phi-\phi_{\rm min}$, we obtain 
\bea \nn
\frac{\delta (\Omega_{\rm DM}/\Omega_{\rm B})}{(\Omega_{\rm DM}/\Omega_{\rm B})} \approx 20\% \l \frac{\delta \phi(t)}{\delta \phi(t_f) } \r \approx 20\% \l \frac{\tf}{t} \r^{3/8}.
\eea
Eventually when all of the constraints are put in place in Fig.~\ref{fig:ain_af_range} , we will find that $1 \gtrsim t_f/t \gtrsim 10^{-2}$. 
As a result, our prediction for $\Omega_{\rm DM}/\Omega_{\rm B}$ will vary between $1-10\%$.
\\

As we estimate below, $\phi$ can very easily overclose the universe.  $\phi$ settles to its new minima in a number conserving manner as long as $\dot m_\phi/m^2_\phi \sim H/m_\phi \lesssim 1$. 
This condition holds for any $t\gtrsim \tf$. 
Thus we obtain the number density of $\phi$ at any time $t\gtrsim \tf$ as 
\bea \label{Eq: nphi}
n_\phi(t) &\lesssim & \frac{\rho_{\rm B}(\tf)}{m_\phi(\tf)}\sqrt{\frac{F}{f_\phi/c_B}} \left(\frac{\af}{R}\right)^3\,. 
\eea
We arrive at this result as follows.  
We first take $V'(\phi)+ V_{\rm scan}'(\phi)=0$ right at the end of relaxation $t_f$.
In the worst case scenario, all of the energy $\rho_{\rm B}(t_f)$ goes straight into producing $\phi$ particles right at the end of relaxation. 
The existence of a minimum $\Lambda^4_{\phi}(t)/F\simeq \rho_{\rm B}(t)/(\fphi/c_B)$ implies that the mass around the $\Lambda$ supported minimum is larger than the mass given by the baryons, giving the $\sqrt{F/(\fphi/c_B)}$ suppression seen in Eq.~\ref{Eq: nphi}.

In the new minima, the mass of $\phi$ is given as $m_\phi=\Lambda_\phi^2/F$. 
With this, we obtain its energy density at $t\gtrsim t_{\rm scan}$ as 
\bea
\!\!\!\!\!\!\!\!\!\!\!\!\!\!\!\!
\rho_\phi(t) = \frac{\Lambda_\phi^2}{(F\fphi/c_B)^{\frac 1{2}}} \frac{\rho_{\rm B}(\trh)}{H_{\rm rh}} \left(\frac{\trh}{\tf}\right)^{\frac{1}{2}}\left(\frac{\af}{R}\right)^3\,. 
\label{eq:rho_phi_after_ts}
\eea 
Thus, once the scanner potential turns on $\phi$ red-shifts as matter with a mass that is independent of the ambient baryon and/or DM energy density.   
\\

After the decay of the reheaton field, we want the universe to be dominated by the SM bath. 
However as $\phi$ redshifts as matter with an energy density much larger than the DM, there is a possibility of $\phi$ domination at the reheating. 
Thus for the universe not to enter a $\phi$ dominated era, we consider $\phi$ decays to $U(1)_D$ gauge bosons $X_{\mu}$. 
We will be considering fermions charged under $SU(D)$ and $U(1)_D$ with mass near the confinement scale.
The effect of these fermions is to generate a $\phi$ coupling to the dark photons. 
Below the $SU(D)$ confinement scale,  
$\phi$ coupling to the dark photon can be written as,
\bea
\mathcal{L}\supset  \frac{r_X}{4}\frac{\phi}{F}X_{\mu\nu}\tilde X^{\mu\nu}\,,
\eea
where $F$ is the effective coupling, $m_\phi = \Lambda_\phi^2/F$, and $r_X$ encodes information about $U(1)_D$ charge and possible loop factors. 
In the presence of the above interaction, $\phi$ decays to dark photon as, 
\bea
\Gamma_{\phi\to XX}=  \frac{r_X^2}{64\pi} \frac{\Lambda_\phi^6}{F^5}\,.
\eea 
Due to the decay of $\phi$, the dark photons energy density can be written as,
\bea
\rho_{X,{\rm decay}}(t\gtrsim \tphi) = \rho_\phi(\tphi)\left(\frac{\aphi}{R}\right)^4\,,
\eea
where, $\tphi$ is defined as  $H(\tphi) = \Gamma_{\phi\to XX}$ and we obtain 
\bea
\frac{\tphi}{\trh} = \frac{64\pi}{r_X^2}\frac{F^5\,H_{\rm rh}}{\Lambda_\phi^6} \,.
\label{eq:phi_XX}
\eea
In order for the $SU(D)$ scanner sector to avoid overclosing the universe, it will also be coupled to the abelian gauge boson $X$. 
The total energy in this sector will come from both the thermal $SU(D)$ sector temperature and from the decay of $\phi$, which occurs close to reheating.  Depending on the parameters, one or the other can be the dominant source of temperature in that sector.  In what follows, we will separate out the energy density in the dark sector into its two components, the decay piece $\rho_{X,{\rm decay}}$ and the pre-existing thermal $D$ sector temperature $\rho_D$. 
We will require both of the pieces to satisfy $N_{\rm eff}$ constraints.
\\

The presence of dark photon in the $SU(D)$ sector leads to extra radiation energy density in the late universe. 
Any extra contribution to the SM photon energy density is usually parameterized in terms of the effective number of neutrino species $N_{\rm eff}$. 
For SM temperature $\TSM\lesssim m_e$ (after the annihilation of the electrons and positrons), the energy density of any additional relativistic species can be recast in terms of its contribution $\Delta N_{\rm eff}$ to $N_{\rm eff}$ as, 
\bea
\rho_{X \, {\rm or}  \, D}(\TSM)= \frac{7\pi^2}{120}\Delta N_{\rm eff} \left(\frac{4}{11}\right)^{4/3}\TSM^4\,.
\eea 
Using the current $95\%$ C.L. limit $N_{\rm eff} = 3.27 \pm 0.29$~\cite{Planck:2018vyg}, with $N_{\rm eff}^{\rm SM}= 3.046$ in the SM, we obtain a constraints on the dark sector energy density at $\Trh$ as,
\bea
\rho_D(\Trh)\lesssim 0.3 \, \Trh^4\,.
\label{eq:delta_Neff}
\eea
Also, the decay of $\phi$ to the dark photons heats up the scanner bath, and leads to an additional constraint of  
\bea
\rho_\phi(\tphi)\left(\frac{\aphi}{\arh}\right)^4 \lesssim 0.3 \,\Trh^4\,,
\label{eq:rho_neff_before_rh}
\eea
if the decay happens before reheating. 
The constraint changes to  
\bea
\rho_\phi(\tphi) \lesssim  0.4 \,\rhoSM(\Trh) \left(\frac{\arh}{\aphi}\right)^4\,,
\label{eq:eho_neff_after_rh}
\eea
if it happens after reheating. 
Note that $\rho_\phi$ can be obtained using Eq.~\eqref{eq:rho_phi_after_ts}. 
\\ 

As $T_{\rm scan}(t_{\rm scan})\simeq\Lambda_\phi$, from Eq.~\eqref{eq:delta_Neff}, we obtain a constraint on $\Lambda_\phi$ as
\bea
\Lambda_\phi\lesssim 12\GeV \left(\frac{2.5\times 10^{-7}}{\tf/\trh}\right)^{1/2}\left(\frac{1}{r_{sf}}\right)\,,
\label{eq:Lambda0_max}
\eea 
for $g_{*S}(\ascan)=15$, and we define $r_{sf}=(t_{\rm scan}/\tf)^{1/2}$ which parameterize the time between the end of relaxation, and scanner sector phase transition. 
$\phi$ can decay to dark photons before or after reheating. 
The former scenario leads to weaker constraints, and that is why we consider it here. 
By combining  Eq.~\eqref{eq:fphi_af}, Eq.~\eqref{eq:rho_neff_before_rh}, and Eq.~\eqref{eq:phi_XX}, we obtain a lower bound on $\Lambda_\phi$ as
\bea
\!\!\!\!\!\!\!\!\!\!\!\!\!\!\!
\Lambda_\phi \gtrsim 4.2\GeV \!\left(\!\frac{F}{10\TeV}\!\right)^2 \!\!
\left(\!\frac{\tf/\trh}{2.5\times 10^{-7}}\!\right)^{\frac{7}{8}}\!\! \left(\!\frac{1}{r_X}\!\right)\!. 
\eea
From above two equations, we obtain an upper bound on $F$ as,
\bea
\!\!\!\!\!\!\!\!\!\!\!\!
F\lesssim 27\TeV\left(\frac{2.5\times 10^{-7}}{\tf/\trh}\right)^{\frac{11}{16}}\!\left(\frac{r_X}{1}\right)^{\frac{1}{2}}\!\left(\frac{1}{r_{sf}}\right)^{\frac{1}{2}}\!.
\eea

Note that $\phi$ can thermalize with the $SU(D)$ plasma through the coupling with $G_D$, if $H\lesssim \Gamma_{\rm th}\simeq \alpha_D^3T_{\rm scan}^3/\pi^2F^2$ where $\alpha_D\equiv g_D^2/(4\pi)$~\cite{Masso:2002np}. 
And if it thermalizes, the dark gauge bosons induce a thermal friction as $\Gamma_{\rm TF}\simeq (D\alpha_D)^5 T^3_{\rm scan}/F^2$~\cite{Moore:2010jd}. 
As $\Gamma_{\rm th},\Gamma_{\rm TF}\propto T_{\rm scan}^3$, these constraints are important at the beginning of the relaxation. 
Thus, if thermalization happens, we need to require $\Gamma_{\rm TF}(\tin)\lesssim H(\tin)$ for successful relaxation. 
However, as both of these requirements have same temperature dependence, we see that as long as the gauge coupling is small, the thermal friction is never important. 
For example, for $\Lambda_\phi\simeq 5\GeV$ and $\mB(\tin)=3\TeV$, $\phi$ thermalizes for 
\bea
\!\!\!\!\!\!\!\!\!\!\!
\!\!
F \lesssim 26\TeV \left(\frac{g_D}{0.05}\right)^3\!\!\left(\frac{\tf/\trh}{2.5\times 10^{-7}}\!\right)^{\frac{1}{2}}\!\left(\frac{r_{sf}}{1}\right)^{\frac{3}{2}}\!,
\eea
whereas the thermal friction is subdominant compare to Hubble as long as
\bea
\!\!\!\!\!\!\!\!\!\!\!\!\!\!
F \gtrsim 800\GeV \left(\frac{g_D}{0.05}\right)^5\!\!\left(\frac{\tf/\trh}{2.5\times 10^{-7}}\right)^{\frac{1}{2}}\!\!\left(\frac{r_{sf}}{1}\right)^{\frac{3}{2}}\!.
\label{eq:F_thermalise}
\eea
for $D=3$. 
\\

Once the scanner sector confines, it would generate glueballs ($\Sigma$) with mass of the order of $\Lambda_\phi$. 
If stable, these glueballs can overclose the universe. 
The glueballs to decay due to the effective coupling between the dark gluons and dark photons $\alpha_D \alpha_X G^2_DX^2/M_{D}^4$ where $\alpha_X$ is the dark fine structure constant and $M_{D} \sim \Lambda_\phi$ is the mass of the vector like fermion charged under $SU(D)\times U(1)_D$.  This fermion was introduced just to allow for the glueballs to decay, similar to how they decay in the standard model. 
Upon confinement, an interaction of the form of $\Sigma F_D^2/f_\Sigma$ would be generated with $f_\Sigma = M_{D}^4/( \alpha_X \Lambda_\phi^3)$. 
The glueballs decay to the dark photons as,
\bea
\!\!\!\!\!\!\!\!\!\!\!\!\!\!\!
\Gamma_{\Sigma\to XX} \simeq \frac{\Lambda_\phi^3}{8\pi f_\Sigma^2} \! \sim 10^{13}\Hrh \left(\frac{\alpha_X}{10^{-4}}\right)^2\left(\frac{\Lambda_\phi}{M_D}\right)^8\!\!\!,
\label{eq:glueball_decay}
\eea 
for $\Lambda_\phi=5\GeV$. 
As, $t_{\rm scan}\gtrsim \tf$, and $(t_f)_{\rm max}/\trh\sim 10^{-6}$ (see Fig.~\ref{fig:ain_af_range}), the glueballs decay as soon as they are formed at $t=t_{\rm scan}$. 
The decay of glueballs to dark photons leads to $\Delta N_{\rm eff}$ constraints which we already discuss in Eq.~\eqref{eq:Lambda0_max}.  
\\

We take $F\sim 25\TeV$, and 
finally, with these parameters we obtain the mass of $\phi$ as 
$m_\phi = \Lambda_\phi^2/F\sim \MeV$. As the zero temperature mass of $\phi$ is larger than $\sim \eV$, the special $\phi$ minimum is not displaced by dense objects like neutron stars. 
At the same time, the scanner is not ruled out by the equivalence principle tests as well, as these tests yield weaker bound in this mass range~\cite{Antypas:2022asj}.

\section{Constraints}\label{sec:constarints}

In this section, we list all the relevant constraints on our model. 
We start our discussion with the temperature of the SM. 

\subsection{Temperature of the SM }\label{sec:TSM}

In this section, we comment on the temperature of the SM sector. 
At the end of inflation, the inflaton predominantly decays to $\Phi_E$. 
Even if the SM is not populated by the inflaton decay, it does acquire a temperature due to the early decay of reheaton. 
At the same time, at the end of baryogenesis, the AD field decays to the SM, and heats up the SM plasma. 
The temperature of the SM can be obtained by solving the evolution equation as,
\bea
\frac{d}{dt}\rhoSM+4 H\rhoSM &=& \Gamma_{\Phi_E}\rho_{\Phi_E} + \Gamma_{\Phi}\rho_{\rm AD}\,,
\eea
where, $\rho_{\Phi_E}$ is the energy density of the reheaton field, and $\rhoSM$ denotes the energy density of the SM relativistic degrees of freedoms. 
The general expression for $\rhoSM$ can be written as,
\bea
\!\!\!\!\!\!\!\!\!\!\!\!
\rhoSM(R) =\frac{1}{R^4}\int^{R} \frac{dR'}{HR'}\, R'^4\left[ \Gamma_{\Phi_E}\rho_{\Phi_E} + \Gamma_\Phi \rho_{\rm AD} \right]\,.
\eea
\\

We solve the above equation in two regimes. 
The full SM temperature follows from the sum of the reheaton and AD source terms. The two results below are limiting expressions for the two individual contributions. In practice, at a given time the larger contribution determines $T_{\rm SM}$. 
If the reheaton dominates in the above equation, then the SM temperature can be obtained as, 
\bea
\!\!\!\!\!\!\!\!\!\!\!\!
\rhoSM\simeq \frac{1}{3}\frac{\Gamma_{\Phi_E} \rho_{\Phi_E}}{H}\propto R^{-1}\,,\,\TSM\sim \rhoSM^{1/4}\propto R^{-1/4}\,,
\eea
where we use  $\rho_{\Phi_E}\propto R^{-4}$, and $\Gamma_{\Phi_E}\propto R$. 
The SM temperature is 
\bea
\!\!\!\!\!\!\!\!\!\!\!
\TSM(\tf) \sim 0.07 \GeV \left(\frac{2.5 \times 10^{-7}}{\tf/\trh}\right)^{1/8}, 
\eea
at the end of the relaxation, which goes upto 
\bea
\!\!\!\!\!\!\!\!\!\!\!
\TSM(\tin) \sim 0.37 \GeV \left(\frac{\mB(\tin)}{\TeV}\frac{5 \times 10^{-4}}{(\tf/\trh)^{1/2}}\right)^{1/4}\!\!, 
\eea
when the relaxation begins. 
Thus, the baryons are non relativistic at the beginning of the relaxation if the SM gets temperature only from early decays. 
At the end of the relaxation, pions are relativistic for
\bea
\frac{\tf}{\trh}\lesssim 4\times 10^{-10}\,,
\label{eq:rel_pion_ed}
\eea
and in that case the pion induced potential is much larger than $\rho_{\rm B}$. 
Even the non relativistic pions can displace $\phi$ from its special minima. 
The pion contribution to $V(\phi)$ can be written as, 
\bea
\!\!\!\!\!\!\!\!\!\!\!\!\!
V_\pi(\tf) = \left(\frac{m_\pi\TSM(\tf)}{2\pi}\right)^{3/2}\TSM(\tf) \, e^{-m_\pi/\TSM(\tf)}\,.
\label{eq:pion_NR}
\eea
Requiring the pion contribution to be sub-dominant compare to that of the baryon, and dark matter, i.e. $V'_\pi(\tf)\lesssim (\rho_{\rm B})'(\tf)$, we obtain,
\bea
\frac{\tf}{\trh}\lesssim 2\times 10^{-6}\,.
\label{eq:af_afmax}
\eea
The above condition constrains the time when the relaxation needs to end irrespective of the initial baryon mass. 
\\

To calculate the SM temperature from the decay of the AD field, we note that, 
$\rho_{\rm AD}\propto R^{-3}$ for $\tbeg \lesssim t\lesssim \tend$. Also, the AD field decays when $H(\tend)=\Gamma_\Phi$. 
Thus we get, 
\bea
\rhoSM(R) &=& \frac{\Gamma_{\Phi}}{R^4}\int^{\aend}_{\abeg} \frac{dR'}{H} R'^3   \rho_{\rm AD}(R')\nn\\
&\simeq& \rho_{\rm AD}(\tend)\left(\frac{\aend}{R}\right)^4\,, 
\label{eq:rhoSm_AD}
\eea
for, $\tbeg\ll \tend$. So, the SM temperature obtained from the decay of the AD scales as $\TSM\propto R^{-1}$. 
Thus, if baryogenesis ends sufficiently early i.e. $\tend\ll\tin$, the SM gets enough time cool down and the baryons become non relativistic when the relaxation starts. 
For $\rho_{\rm AD}(\tbeg)= 10^{30}\GeV^4 $, we find the temperature of the SM at the end of baryogenesis as 
\bea
T_{\rm SM}(\tend)\sim 13\TeV\,, 
\label{eq:tsm_end}
\eea
for $\tbeg/\tend=10^{-8}$. We use $g_*(T)=106.7$ in the above estimate. 
We want to know the temperature of the SM at the beginning of the relaxation. 
Using Eq.~\eqref{eq:rhoSm_AD} we obtain 
\bea
\!\!\!\!\!\!
\rhoSM(\tin)= \rho_{\rm AD}(\tbeg)\left(\frac{\abeg}{\aend}\right)^3\left(\frac{\aend}{\ain}\right)^4\,,
\eea
and, 
\bea
\rhoSM(\tin)\simeq \left(2 \TSM(\tin)\right)^4\,,
\eea
where we use $g_*\sim 50$ as the SM relativistic degrees
of freedoms after the QCD confinement. 
Thus we obtain $\TSM(\tin)\simeq \GeV$ for $\tend/\tin=10^{-8}$. 
\\

For the relaxation to proceed as described in Section~\ref{sec:phi_dynamics}, the baryons need to be non relativistic at the beginning of the relaxation. 
Thus by requiring $\TSM(\tin)\lesssim \mB(\tin)$, we get a constraint on the ratio of time between the end of baryogenesis and the beginning of relaxation as, 
\bea
\frac{\tin}{\tend}
\gtrsim 
2.5\times 10^2  \left(\frac{\TeV}{\mB(\tin)}\right)^2\,,
\label{eq:baryo_NR_baryon}
\eea
for $\rho_{\rm AD}(\tbeg)=10^{30}\GeV^4$, and $\tbeg/\tend=10^{-8}$. 
Even when the baryons are non relativistic, the pions can still be relativistic, and can dominate the evolution of $\phi$ at the beginning of the relaxation. 
Requiring the pion contribution to the $\phi$ potential is sub-dominant compare to that of the baryons at $t=\tin$ i.e. $m^2_\pi\TSM^2(\tin)/8 \lesssim \rho_{\rm B}(\tin)$, using  Eq.~\eqref{eq:eta_S}, we obtain 
\bea
\!\!\!\!\!\!\!\!\!\!\!\!\! 
\frac{\tin}{\trh}\lesssim 
3.6\times 10^{-5} \left(\frac{\rho_{\rm AD}(\tbeg)}{10^{30}\GeV^4}\right)\!\! \left(\frac{100\TeV}{m_\Phi}\right)^3, 
\label{eq:baryo_pi_vb}
\eea
for $\tbeg/\tend=10^{-8}$, which is a much weaker requirement than Eq.~\eqref{eq:af_afmax}. 
The above equation is conservative as the pions are relativistic only for 
$\tin/\trh\lesssim 2 \times 10^{-9}(\GeV/\mB(\tin))$. 
If the pions are non-relativistic at the beginning of the relaxation,  their contribution is exponentially small and thus can be safely neglected. 
Thus the requirement of the pion contribution to the $\phi$ potential needs to be subdominant compare to that of the baryons at the beginning of the relaxation does not lead to any new constraints.   

We also require that the pions to be non relativistic at the end of relaxation i.e.  $m_\pi(\tf)\gtrsim T_{\gamma}(\tf)$, and obtain a constraint as
\bea
\!\!\!\!\!\!\!\!\!\!\!\!\!\!\!
\frac{\tf}{\trh}\gtrsim 2\times 10^{-9} \left(\frac{\rho_{\rm AD}(\tbeg)}{10^{30}\GeV^4}\right)^{1/2}\!\! \left(\frac{100\TeV}{m_\Phi}\right).
\label{eq:rel_pion_AD}
\eea
This is similar to the case discussed in Eq.~\eqref{eq:rel_pion_ed}; however the SM temperature is dominated by the decay of the AD field here. 
However, the above constraint is weaker as the relativistic pion contribution to the $\phi$ potential is sub-dominant compare to that of the baryons at $t=\tf$ as long as $\tf/\trh\lesssim 3.6\times 10^{-5}$.

\subsection{Summary of the Constraints}

Our model has a lot of moving parts, but not so many constraints. 
So here we summarize many of the different considerations that need to be satisfied for the model to function. 

\begin{enumerate}

\item Effective field theory constraint on the axion  decay constant: $f_a\lesssim f_\phi$. 
Discussed in Eq.~\eqref{eq:fa_less_fphi}. 
As $f_a$ decreases during relaxation, this requirement needs to be satisfied at the beginning of the relaxation i.e. $f_a(\tin)\lesssim f_\phi$. Discussed in Eq.~\eqref{eq:fa_fphi_final}. \label{item_1}

\item Baryons are non relativistic at the beginning of the relaxation i.e. $\mB(\tin)\gtrsim \TSM(\tin)$. Discussed in Eq.~\eqref{eq:baryo_NR_baryon}. \label{item_2}

\item \label{item_3}The potential of $\phi$ is dominated by the baryons during the relaxation. 
Most importantly pion contribution is subdominant at $\tin$ i.e. $m^2_\pi\TSM^2(\tin)/8 \lesssim \rho_{\rm B}(\tin)$. 
Discussed in Eq.~\eqref{eq:baryo_pi_vb}. 

\item \label{item_4}Relaxation starts when Hubble is equal to the mass of the scanner field $\phi$ i.e. $H=m_\phi$. Discussed in Eq.~\eqref{eq:mphi_hti}. 
This fixes $f_\phi$ as a function of $\tf$. 
Discussed in Eq.~\eqref{eq:fphi_af}. 

\item\label{item_5} Initial axion misalignment angle needs to be $\theta_0\lesssim 2\pi$ since the axion starts to oscillate to the end of the relaxation. We require the requirement to hold at $t=\tf$. Discussed in Eq.~\eqref{eq:theta_2Pi}.

\item\label{item_6} The pion contribution to $V(\phi)$ needs to be sub-dominant compare to baryons at the end of relaxation.  
Discussed in Eq.~\eqref{eq:rel_pion_ed}, Eq.~\eqref{eq:rel_pion_AD}, and Eq.~\eqref{eq:af_afmax}. 

\item\label{item_7} Several additional constraints such as $\Delta N_{\rm eff}$ (Eq.~\eqref{eq:delta_Neff}), thermal friction of $\phi$ (Eq.~\eqref{eq:F_thermalise}), and/or overclosing the universe by stable particles (Eq.~\eqref{eq:glueball_decay}) arise due to the temperature dependence of the scanner potential. 
These constraints can be satisfied by choosing appropriate parameters of the scanner sector as discussed in Section~\ref{sec:scanner_potential}, and thus are not discussed here.  

\item\label{item_8} The axion thermalization at reheating leads to a constraint on $f_a$ as
\bea
f_a \gtrsim \alpha^{3/2} \sqrt{\Trh \Mpl} \sim 10^5 \GeV\,, 
\label{eq:fa_thermalisation}
\eea
for $\Trh\sim 10\MeV$. 
\end{enumerate}

\begin{figure}
    \centering
    \includegraphics[width=\columnwidth]{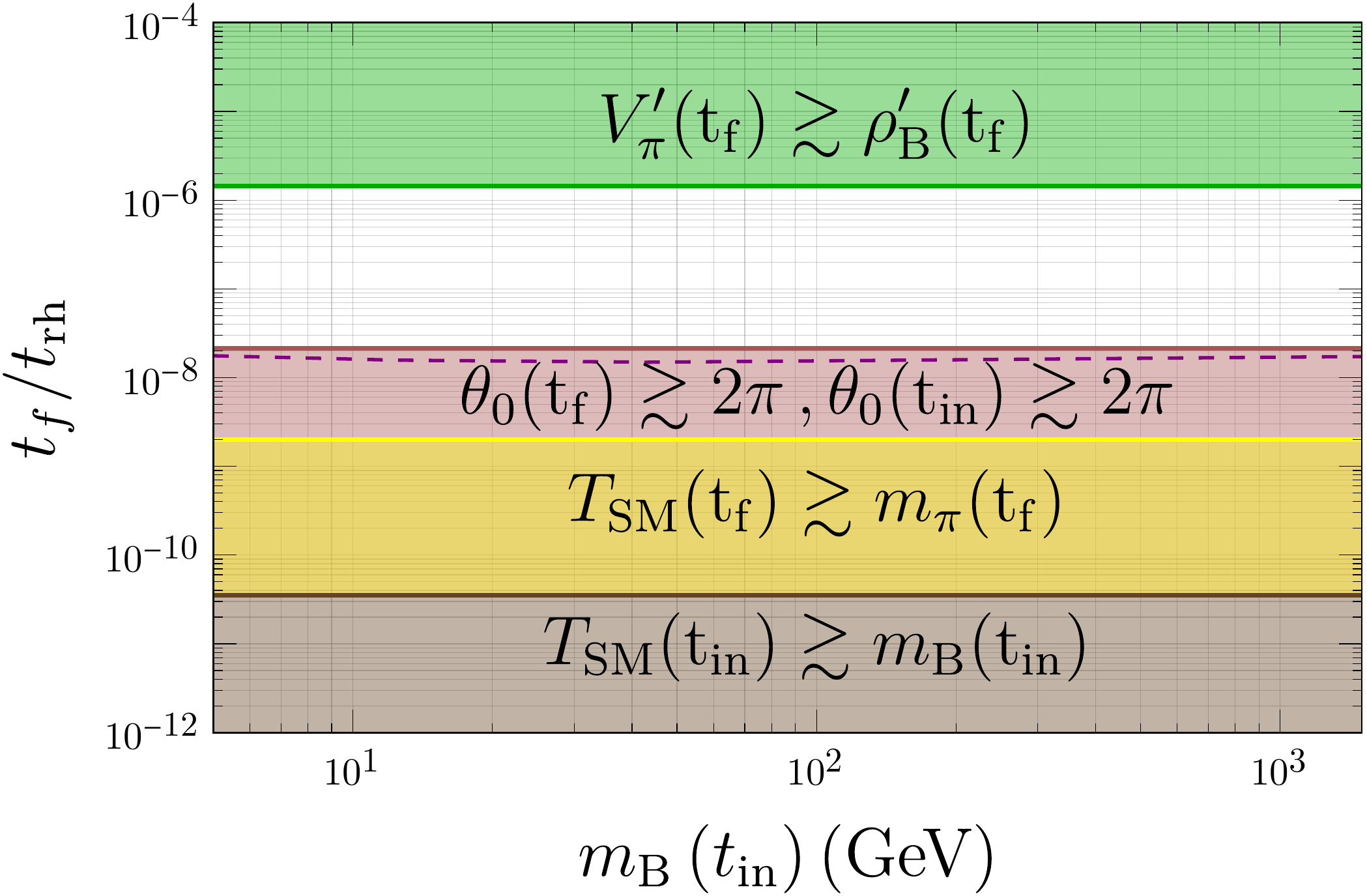}
    \caption{Plot of the time when relaxation ends $(\tf)$ normalized to the time of reheating $(\trh)$ as a function of the baryon mass at the beginning of the relaxation $\mB(\tin)$. 
    The green, and the yellow shaded regions are excluded by the requirement that the pion contributions to the $\phi$ potential at the end of relaxation need to be subdominant compare to that of the baryons. 
    In the green shaded region, pions are non relativistic at $\tf$, whereas in the yellow shaded region they are relativistic.
    The overlapping magenta, and pink shaded regions are excluded by the requirement that the axion misalignment angle is less than $2\pi$ through out the relaxation phase. 
    The brown shaded region is excluded by the requirement that at the beginning of the relaxation baryons are non relativistic. 
    }
    \label{fig:ain_af_range}
\end{figure}

In Fig.~\ref{fig:ain_af_range}, we plot various constraints discussed above. 
These restrict the time when relaxation ends as a function of duration of the relaxation phase which is dictated by the initial baryon mass. 
At the end of relaxation, thermal correction to $\phi$ potential from pions can spoil the relaxation by driving $\phi$ away from its special minima as discussed in Eq.~\eqref{eq:rel_pion_ed} and Eq.~\eqref{eq:pion_NR}({\color{red}item~\ref{item_6}} in the summary list). 
The green shaded region is excluded due to the non-relativistic pion contribution being dominant over the baryon contribution. 
As can be see from the figure, this constraint provides the upper bound on $\tf$. 
The requirement that pions are non-relativistic at the end of relaxation sets a lower bound on $\tf$, as indicated by the yellow band. 
A stricter lower bound on $\tf$ is obtained by requiring that the initial misalignment angle of the QCD axion is smaller than $2\pi$. 
We thus require  $\theta_0(\tf)\lesssim 2\pi$ as discussed in Eq.~\eqref{eq:theta_2Pi} and item~\ref{item_5} in the summary list. 
This constraint is shown by the magenta shaded region in Fig.~\ref{fig:ain_af_range}. 
We depict $\theta_0(\tin)\lesssim 2\pi$ constraint by the dashed purple line. 
As the value of $\theta_0$ barely changes during relaxation, these two regions overlap significantly. 
Also, instead of the misalignment angle at $\tin$, one should consider its value at $\tosc$ and this leads to an even weaker requirement.  

Bounds from baryons being relativistic at the beginning of the relaxation is shown by the brown shaded region in Fig.~\ref{fig:ain_af_range} (item~\ref{item_2} in the summary list). 
The relativistic pions contribution to $V(\phi)$ becoming dominant over the baryons at $\tin$ (item~\ref{item_3} in the summary list) yields a much weaker constraint, and thus are not shown here. 
As discussed in Eqs.~(\ref{eq:baryo_NR_baryon} -\ref{eq:baryo_pi_vb}),
these bounds depend on when baryogenesis ends and the relaxation starts, and thus can be satisfied easily if these two timescales are well separated. 
We depict the bounds here for our benchmark parameters for $\tend$ (see baryogenesis section for more details). 
\\

Thus, the maximum allowed $\tf$ is set by the requirement that the pion contribution to the $\phi$ potential do not displace it from the special minimum (c.f. Eq.~\eqref{eq:af_afmax}), whereas the minimum allowed $\tf$ is fixed by the requirement that the axion misalignment angle is less than $2\pi$ (c.f. Eq.~\eqref{eq:theta_2Pi}).
\\

\begin{figure}
    \centering
    \includegraphics[width=\columnwidth]{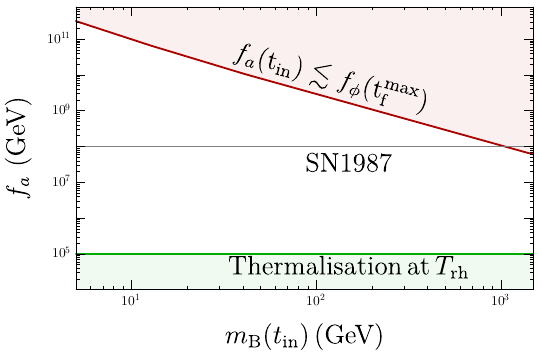}
    \caption{Plot of axion decay constant at the end of relaxation as a function of initial baryon mass.  
    The red line shows the maximum axion decay constant that can be obtained in our construction. This is obtained by using Eq.~\eqref{eq:fa_fphi_final} and Eq.~\eqref{eq:af_afmax}.  
    In the green shaded region axion thermalizes at the time of reheating.  
    The gray line depicts the bound from supernovae 1987A~\cite{Raffelt:2006cw}. 
    }
    \label{fig:fa_af_r}
\end{figure}

After obtaining the minimum and maximum allowed $\tf$, we plot the allowed axion decay constant as a function of initial baryon mass in Fig.~\ref{fig:fa_af_r}. 
The maximum allowed axion decay constant is fixed by the effective theory requirement of $f_a(\tin)\lesssim f_\phi$ (item~\ref{item_1} in the summary list). 
This constraint is proportional to $\tf^{1/4}$ as can be seen from Eq.~\eqref{eq:fa_fphi_final}. 
Starting from Eq.~\eqref{eq:fa_fphi_final}, we obtain maximum axion decay constant at the end of the relaxation by using either Eq.~\eqref{eq:fa_in_mBin_less_mt}, Eq.~\eqref{eq:fa_in_mBin_less_mb} or Eq.~\eqref{eq:fa_in_mBin_less_mc} for maximum $\tf$. 
In Fig.~\ref{fig:fa_af_r}, we show this constraint by the red shaded region. The lower limit on $f_a$ comes from the consideration that the axion can thermalize with the SM plasma at the time of reheating as discussed in Eq.~\eqref{eq:fa_thermalisation} (item~\ref{item_8} in the summary list). 
We show this constraint by the green shaded region in Fig.~\ref{fig:fa_af_r}. 
The bound from SN1987A cooling~\cite{Raffelt:2006cw} is shown by a gray line in Fig.~\ref{fig:fa_af_r}, despite this being subjected to recent critical study~\cite{Bar:2019ifz}. 
Neutron star cooling provides additional constraint on axion-nucleon coupling~\cite{PhysRevLett.53.1198,Buschmann:2021juv}, which can be translated to a bound on $f_a\lesssim 10^{8}\GeV$ with some model dependence.

\begin{figure}
    \centering
    \includegraphics[width=\columnwidth]{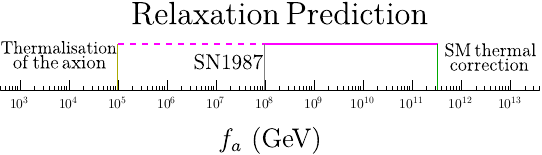}
    \caption{A plot of the axion decay constant that can be accommodated by our relaxation mechanism.}
    \label{fig:fa_range}
\end{figure}

Therefore, we obtain the range of axion decay constant to be
\bea
10^{8}\GeV\lesssim f_a\lesssim 3.2\times 10^{11}\GeV\,,
\eea
where the coincidence problem can be addressed along with a composite QCD axion DM as shown in  Fig.~\ref{fig:fa_range}.   

\section{Summary and conclusion}
\label{sec:conclusion}

In this article, we explored the possibility that a new relaxation mechanism could explain the dark matter - baryon coincidence problem in the context of QCD axion dark matter. 
The relaxation mechanism solving this problem requires a coupling to both dark matter and QCD.
The QCD axion is a dark matter candidate that is intrinsically tied to QCD.  As such, it is natural that anything that modifies the QCD axion also modifies QCD.  
In fact, this connection is so strong that it is a bit surprising that the model has enough flexibility to accommodate current data.  

By scanning the axion decay constant, $f_a$, the axion and baryon masses are connected with only integer degrees of freedom, a ratio of beta functions.  The integer degree of freedom is required to reproduce the observed baryon to dark matter ratio to within 1-10\% depending on if the new physics modifies the cosmological fit. 
The fact that there is a solution is amazing.  Amusingly this entire approach can be excluded in several very non-standard ways. 
A precision measurement of $\rho_{\rm B}/\rho_{\rm DM}$ and a better understanding of how the proton mass depends on the UV QCD gauge coupling are just two of the odd ways in which such a theory could be tested. 

Finally, there is still much to think about and improve on this approach. 
Firstly, the cosmology we consider is complicated and it would be interesting if there were a simpler cosmology that works. 
Secondly, an important assumption, present for all theories where finite density/temperature effects are important, is that the vacuum potential for $\phi$ has been highly suppressed. 
It would be interesting if there was a convincing solution to this puzzle.

\section*{Acknowledgments}

The authors would like to thank Zackaria Chacko and William DeRocco for useful discussions and  comments. 
The authors are supported by the National Science Foundation under grant number PHY2210361 and the Maryland Center for Fundamental
Physics. 
In addition, DB is supported by the Department of Energy under grant number DE-SC0007859.

\newpage
\bibliographystyle{apsrev4-1}
\bibliography{ref.bib}

\appendix

\renewcommand{\theequation}{\thesection.\arabic{equation}}
\setcounter{section}{0}
\renewcommand{\thesection}{\Alph{section}}

\section{Composite QCD axion}\label{app:composite_axion}

In this section, we briefly introduce the composite axion models~\cite{Kim:1984pt,Choi:1985cb,Kaplan:1985dv}. 
In these models, the axion is not a fundamental particle, rather the composite pion-like object of some gauge group $G$. 
Because the axion is a composite, its decay constant is given by the confinement scale of the gauge group $G$, similar to the case of the pion decay constant being related to $\Lambda_{\rm QCD}$~\cite{Witten:1980sp}. 
As the axion decay constant is generated by the physics of dimensional transmutation, the composite axions are more protected against the quality operators compare to that of the DFSZ and/or KSVZ models~\cite{Randall:1992ut,Dobrescu:1996jp}. 
To highlight our purpose, in the following we provide one example of composite axions.  
For simplicity, we take $G=SU(N)$. 
 
Following~\cite{Kim:1984pt}, we consider massless vector-like fermions $\psi$ and $\xi$ that transform under $SU(N)$ and $SU(3)$ respectively as, 
\bea
\psi=(N,3)\,\,{\rm and}\,\,\,\xi= (N,1)\,. 
\label{eq:fermion_composite_axion}
\eea
We assume that $SU(N)$ confines at a scale $\Lambda_N\gg \Lambda_{\rm QCD}$. 
As long as $SU(N)$ confines with chiral symmetry breaking, it will break the flavor symmetries down to the diagonal via   $\left<\bar\psi_{L,i}\psi_{R,j}\right>=\left<\bar\xi_{L,i}\xi_{R,j}\right>= \Lambda_N\,\delta_{ij}$. 
In the limit of QCD coupling goes to zero ($g_s\to 0$), the global symmetry is $SU(4)_L\times SU(4)_R\times U(1)_V$, which is spontaneously broken down to $SU(4)_{L+R}\times U(1)_V$ by the fermion condensates at the scale $\Lambda_N$. However, a small but non zero $g_s$, forces the condensate to preserve color, and the 15 pseudo-Nambu Goldstone Bosons transform as $1\oplus 3\oplus\bar 3\oplus 8$ representations of $SU(3)_C$. 
The color singlet combination of 
\bea
a\sim \left(\bar\psi\gamma^5\psi-3\bar\xi\gamma^5\xi\right)\,,
\eea 
is identified with the axion.

When QCD is turned on (i.e. $g_s\neq 0$), the $SU(4)_{L+R}$ is broken explicitly to the gauged $SU(3)_C$ and a global $U(1)$.  This gauging gives a mass $\sim g_s \Lambda_N/4 \pi$ to all of the non-axion massless particles.
Meanwhile the axion gets a tiny mass from QCD instantons, 
since the associated current 
\bea
J^\mu_A = \frac{1}{2\sqrt{6}}\left(\bar\psi\gamma^\mu\gamma^5\psi-3\bar\xi\gamma^\mu\gamma^5\xi\right)\,,
\eea
is anomalous under QCD as 
\bea
\partial_\mu J^\mu_A = \frac{g_s^2}{16 \pi^2}\frac{N}{2\sqrt{6}} G\widetilde G\,.
\eea
The axion interaction with the SM can be written as 
\bea
\mathcal{L} \supset \frac{g_s^2}{32 \pi^2}\frac{a}{f_a}G\widetilde G\,,
\eea
with axion decay constant $f_a= \sqrt{6}\Lambda_N/N\,$. 
Note that the simplest composite axion model considered here have domain walls, which can be circumvented by additional model building (see e.g.~\cite{Kim:1984pt} for a discussion). 
In our case, we are considering axion DM production from the misalignment mechanism with a reheating temperature of $\sim \mathcal{O}(10\MeV)$. 
Therefore our conclusion does not require any additional model building regarding this. 

\section{Details of Affleck-Dine benchmark} \label{app:AD_details}

In this appendix we present a full derivation of baryon asymmetry estimate and benchmark decay widths used in Section~\ref{sec:Baryogenesis}. 

\subsection{Asymmetry generation in the AD condensate}
We consider a complex scalar $\Phi$ that carries a global baryon number charge of $Q_\Phi$ with the following potential  
\bea
V(\Phi,\Phi^\dagger)= m_\Phi^2 (\Phi^\dagger\Phi) + \epsilon\,m_\Phi^2 (\Phi^2+\Phi^{\dagger 2})\,,
\eea
where $m_\Phi$ is the mass of $\Phi$, and a small real $\epsilon$ softly breaks the symmetry.

To see how the baryon asymmetry is generated, let us write $\Phi = (\phi_1+i\phi_2)/\sqrt{2}$. 
The EOMs for $\phi_1$, and $\phi_2$ can be written as
\bea
\ddot\phi_i+3H\dot\phi_i+m_i^2\phi_i&=&0\,,
\label{eq:EOM_phi1_phi2}
\eea
for $i=1,2$ and we define $m_{1,2}^2=m_\Phi^2(1\pm2\epsilon)\,$. 
The AD field is frozen for $H\gtrsim m_\Phi$. 
At a time $t=\tbeg$, the fields start to oscillate and the solutions to the EOMs can be written as
\bea
\!\!\!\!\!\!\!\!\!\!\!\!
\phi_{i}(t) \approx \phi_{i}(\tbeg)\left(\frac{\abeg}{R(t)}\right)^{3/2}\!\!\!\cos\left[m_{i}(t-\tbeg)\right],
\eea
where, $\phi_{1,2}(\tbeg)$ are the initial misaligned values of $\phi_1$ and $\phi_2$. 
The onset of the oscillation is obtained by solving $3 H(\tbeg)= m_{1,2}\simeq m_{\Phi}\,$, and we also take $\phi_{1}(\tbeg) \sim \phi_2(\tbeg)= \Phi(\tbeg)$. 
The baryon  asymmetry in the $\Phi$ condensate can be written as, 
\bea
\!\!\!\!\!\!\!\!\!\!\!
n(t) = i\,Q_\Phi \! \left(\Phi^\dagger\dot\Phi-\dot\Phi^\dagger \Phi\right) \! \!= Q_\Phi \!\!\left(\dot\phi_1\phi_2-\phi_1\dot\phi_2\right).
\eea

To transfer the asymmetry from the $\Phi$ condensate to the SM, we consider the  decay of the AD field to the SM with decay width $\Gamma_\Phi$. 
In the presence of the decay term, the time evolution for $n$ changes as, 
\bea
\dot n+(3H+\Gamma_\Phi)n = 4\epsilon\,Q_\Phi m_\Phi^2\phi_1 \phi_2\,. 
\eea
For convenience, we define the co-moving asymmetry as $n_c(t)=(R/\abeg)^3\, n_{\rm SM}(t)$, where $n_{\rm SM}$ is the baryon number density in the Standard Model. 
The total co-moving asymmetry transferred to the SM can be obtained as, 
\bea
\!\!
n_c(t) \!&=& 4\epsilon Q_{\Phi} m_\Phi^2\int_{\tbeg}^{t} \!\!\! dt' \!\! \left(\frac{t'}{\tbeg}\right)^{\frac{3}{2}} \!\!\!\phi_1(t')\phi_2(t') e^{-\Gamma_\Phi(t'-\tbeg)}\nn\\
&=& 4\epsilon Q_{\Phi} m_\Phi^2 \Phi^2(\tbeg)\times \mathcal{I}(t)\,, 
\eea
where, we define 
\bea
\!\!\!\!\!
\mathcal{I}(t) &=& 
\int_{\tbeg}^{t}  dt' \cos\left[m_1(t'-\tbeg)\right] \cos\left[m_{2}(t'-\tbeg)\right]\nn\\
&& \,\,\,\,\,\,\,\,\, \times \,\, e^{-\Gamma_\Phi(t'-\tbeg)} \nn\\
&=& \frac{\gamma(2+\gamma^2)}{m_\Phi(4\gamma^2+\gamma^4+16\epsilon^2)} +\mathcal{O}(e^{-\Gamma_\Phi t})\,,
\eea
and $\gamma=\Gamma_\Phi/m_\Phi$. 
Thus at the end of baryogenesis $t=\tend$, we obtain the total  asymmetry transferred to the SM as, 
\bea
\!\!\!\!\!\!
n_{\rm tot}(\tend) &= &\left(\frac{\abeg}{\aend}\right)^3 n_c(\tend)\\
&\simeq& 4 Q_{\Phi}\, n_{\rm AD}(\tbeg) \left(\frac{\abeg}{\aend}\right)^3 \times \mathcal{F}(\epsilon,\gamma)\,\nn,
\eea
where we define  
\bea
\mathcal{F}(\epsilon,\gamma)= \frac{\epsilon\gamma(2+\gamma^2)}{(4\gamma^2+\gamma^4+16\epsilon^2)}\,,
\eea
and $n_{\rm AD}(\tbeg)\simeq m_\Phi \Phi^2(\tbeg)$
is the number density of the AD field when it starts to oscillate.

Note that $\mathcal{F}(\epsilon,\gamma)$ encodes the underlying details of the AD mechanism. 
For a given $n_{\rm AD}$, the maximal baryon asymmetry is achieved when $\mathcal{F}(\gamma,\epsilon)$ is maximised. 
This is obtained 
for $\epsilon \simeq\gamma/2$ and we get $\mathcal{F}_{\rm max}=1/8$ for $\gamma\ll 1$. 
Thus, we obtain maximal baryon asymmetry when the decay width is of order the oscillation time scale as discussed previously. 

\subsection{Estimate of the final co-moving asymmetry}

Using the maximal value of $\mathcal F$, the final co-moving asymmetry can be written as
\bea
\!\!\!\!\!\!\!\!\!\!\!\!\!\!
\eta_s\equiv\left.\frac{n_{\rm tot}}{s}\right|_{{\tend}}
\!\!\!\! \simeq \frac{Q_{\Phi}}{2}
\frac{ n_{\rm AD}(\tbeg)}{s(t_{\rm ref})}
\left(\frac{\abeg}{R_{\rm ref}}\right)^3 
\,,
\label{eq:eta_S}
\eea
where, $s(t_{\rm ref})=5.89\times 10^{-7} \GeV^3$ is the entropy density at a reference time $t_{\rm ref}$ after reheating has completed, with a reference temperature of $T_{\rm ref}=5\MeV$ and a corresponding scale factor of $R_{\rm ref}/\arh = 2$.  
Using $3H(\abeg)=m_{\Phi}$, we finally obtain
\bea
\!\!\!\!\!\!\!\!\!
\eta_s
\simeq 10^{-10}
\left(\frac{ \rho_{\rm AD}(\tbeg)}{10^{30}\GeV^4}\right)\left(\frac{ 100 \TeV}{m_\Phi}\right)^{5/2},
\label{eq:etas_final}
\eea
where $\rho_{\rm AD}(\tbeg)=m_\phi n_{\rm AD}(\tbeg)$ is the initial energy density of the AD field, and we take $Q_\Phi=2$. 
The time when baryogenesis starts can be obtained as,
\bea
\frac{\tbeg}{\trh} = 1.4\times 10^{-27} \left(\frac{100\TeV}{m_\Phi}\right).
\eea 

\subsection{Decay chain and benchmark widths}

To transfer the asymmetry to the Standard Model, we consider
\bea
\!\!\!\!\!\!\!\!\!\!\!\!\!\!
\mathcal{L} \supset y_1 \Phi \psi_R \psi_R+ y_2 \psi_R u_R\, S^\dagger + \epsilon_{pqr}y_3 S^{p}\, d^{q}_R d^{r}_R \,,  
\eea
where $\psi_R$ is a color-singlet fermion with baryon number $-1$, $S$ is a colored scalar with baryon number $-2/3$ and hypercharge $2/3$, and $u_R,d_R$ are SM right-handed quarks. These interactions imply that $\Phi$ carries baryon number $Q_\Phi=2$ and decays through the cascade
\bea
\Phi \to \psi_R\psi_R,\qquad
\psi_R\to u_RS,\qquad
S\to dd\,.
\eea

To simplify the calculation, we take the mass hierarchy to be  $m_\Phi/2\gtrsim m_\psi\gtrsim m_S$. 
For the $\Phi\to \psi_R\psi_R $ decay, the width can be written as, 
\bea
\!\!\!\!\!\!\!\!\!
\Gamma_{\Phi} = \frac{y_1^2\,m_\Phi}{8\pi} \left(1-\frac{4 m_\psi^2}{m_\Phi^2}\right)^{1/2}\left(1-\frac{2 m_\psi^2}{m_\Phi^2}\right)\,.
\eea
For $\psi\to u S$, and $S\to dd$ decay, we consider the SM quarks to be mass-less and obtain
\bea
\!\!\!\!\!\!\!\!\!\!\!\!\!\!\!\!\!\!
\Gamma_\psi = \frac{3 y_2^2\,m_\Psi}{16\pi} \left(1-\frac{m_S^2}{m_\psi^2}\right)^{2}\!\!\!\,{\rm and}\,\,\, \Gamma_S = \frac{y_3^2\,m_S}{8\pi}\,, 
\eea
respectively. 
Note that, for $\Gamma_S$, we consider $S$ decaying into quarks of two different flavors. 
If $\Phi\to\psi\psi$ is the rate determining step in the cascade decay of $\Phi$, we obtain the duration of baryogenesis as
$\tbeg/\tend= 3\gamma$ by using $H(\tend)=\Gamma_\Phi$. 
A hierarchical decay widths of $\Gamma_\Phi\ll \Gamma_\psi\ll \Gamma_S$, can be obtained by choosing appropriate masses and couplings of the fields as discussed in the following. 
\\

Finally, we provide a benchmark point for the our baryogenesis model. 
Note that the exact values of the parameters are not important, as long as the hierarchy is maintained. 
We choose $m_\Phi=100\TeV$, $m_\psi = 45\TeV$, $m_S=40\TeV$, $y_1=3.7\times 10^{-4}$, and $y_2 =y_3=0.1$. 
With these set of parameters, we obtain $\Gamma_\Phi \simeq 0.3 \MeV$, $\Gamma_\psi \simeq 1.2 \GeV$, and $\Gamma_S\simeq 16 \GeV$. 
Thus, $\gamma\simeq 0.33 \times 10^{-8}\sim \epsilon$, and the period of baryogenesis lasts $\tbeg/\tend=10^{-8}$.

\subsection{Washout and phenomenological remarks}

At the end of baryogenesis, when the AD field decays to the SM, the inverse processes can lead to washout of the asymmetry. 
As the baryon number violating spurion $\epsilon$ has baryon charge of $-4$, the leading washout processes would be through $\Phi$ induced $(\psi_R \partial \psi_R)^2$ operator. 
The rate of the washout process can be estimated as  
\bea
\Gamma_{\Delta B=4} \sim \frac{\epsilon^2 y_1^4\TSM^6}{m_\Phi^8}n_\psi\,,
\eea
where $n_\psi$ is the number density of $\psi$ at the end of the baryogenesis. 
In Section~\ref{sec:TSM}, we have calculated 
the SM temperature, and have found that at the end of the Baryogenesis  $\TSM(\tend)\sim 13\TeV\lesssim  m_\psi=45\TeV$. 
Thus $\Gamma_{\Delta B=4} \ll H(\tend)\sim\MeV\,$ i.e. baryon wash out rate is negligible in our case. 
For $\TSM\sim \mathcal{O}(\TeV)$, the electroweak sphaleron processes are still active, and can lead to washout of the asymmetry~\cite{Cohen:1993nk}. 
However these processes can lead to at-most an $\mathcal{O}(1)$ change in the estimate and does not alter the result. 
An effective dimension-18 operator with $12$ quarks can be obtained after integrating out $\Phi,\psi$ and $S$.  Washout from this operator is negligible.
Inside a neutron star, this could induce two neutrons to convert into two antineutrons. 
But despite the neutrons being relativistic inside a neutron star, the rate is highly suppressed, and thus not shown here. 

The existence of new colored states is constrained by LHC observables and precision measurements of flavor-changing neutral currents (FCNC), such as meson-antimeson oscillations. 
Current limits from dijet angular distribution~\cite{CMS:2018ucw} set to a bound on the mass of the colored scalar at $m_S\gtrsim 800\GeV$, which is on par with the LHC constraint on pair-produced colored states    $\sim\TeV$~\cite{ParticleDataGroup:2024cfk}. 
FCNC observables impose stronger constraints on the scale of the four-quark operators that can be generated by integrating out the new states~\cite{Ferrari:2023slj}. 
However, these observables depends on the flavor structure of the new colored states, which are not crucial for baryogenesis, and can be circumvented by appropriately assigning the flavor structure of the Yukawa couplings.

\section{Illustrative Evolution of Energy Densities and Scanned Scales}

\begin{figure}
    \centering
    \includegraphics[width=\columnwidth]{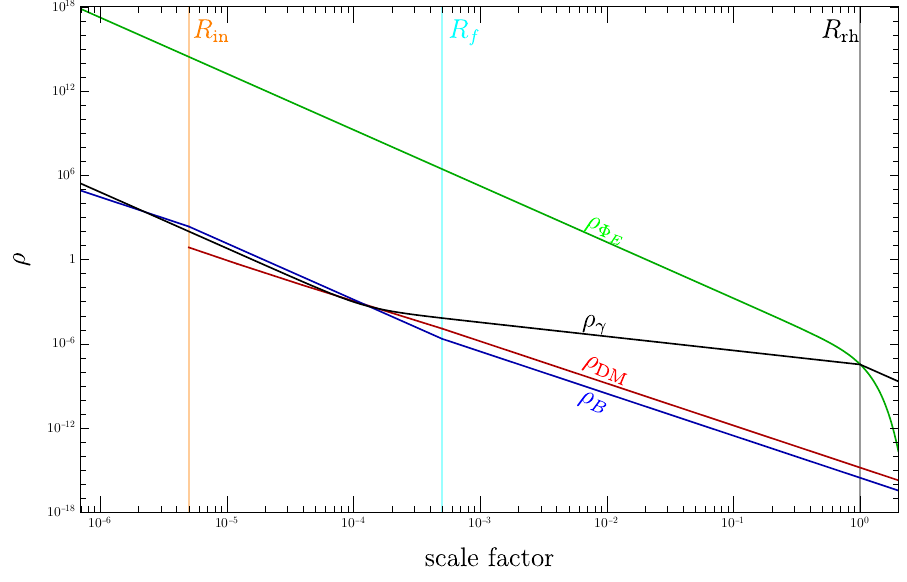}
    \caption{ 
    Energy densities of the reheaton (green), baryon (blue), dark matter (red), and the relativistic part of the SM (black) as a function of scale factor. The vertical orange, turquoise and the gray lines denote the scale factor associated to the era of beginning of the relaxation ($R_{\rm in}$), end of the relaxation ($R_{f}$) and reheating ($R_{\rm rh}$)  respectively. For the purpose of this plot, we consider two decades of relaxation {\it i.e.} $R_{\rm in}/R_f=10^{-2}$ and take the final axion decay constant to be $f_a=10^9\GeV$. With these choices of parameters, the axion starts to oscillate at $R=R_{\rm in}$.  
    }
    \label{fig:app_rho_various_R}
\end{figure}

\begin{figure}
    \centering
    \includegraphics[width=\columnwidth]{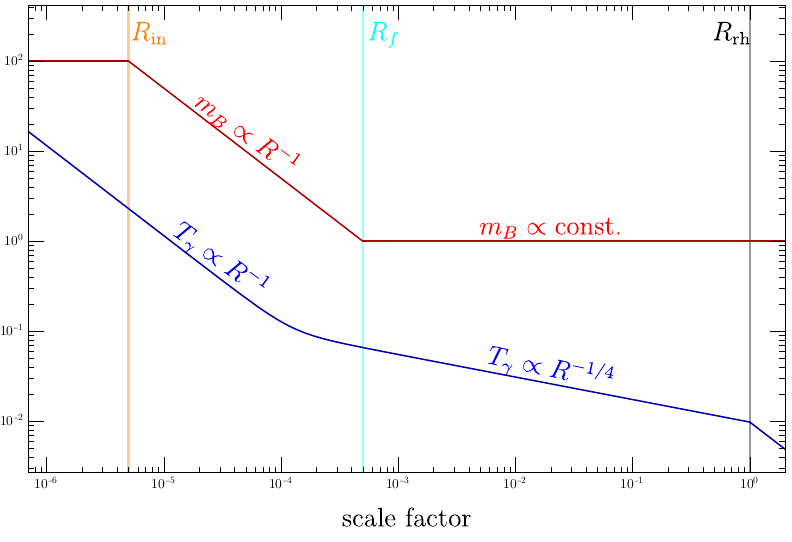}
    \caption{ 
    Plot of the SM temperature (blue) and baryon mass (red) as a function of scale factor. 
    The rest of the figure description matches that of  Fig.~\ref{fig:app_rho_various_R}. 
    }
    \label{fig:app_SM_R}
\end{figure}

\begin{figure}
    \centering
    \includegraphics[width=\columnwidth]{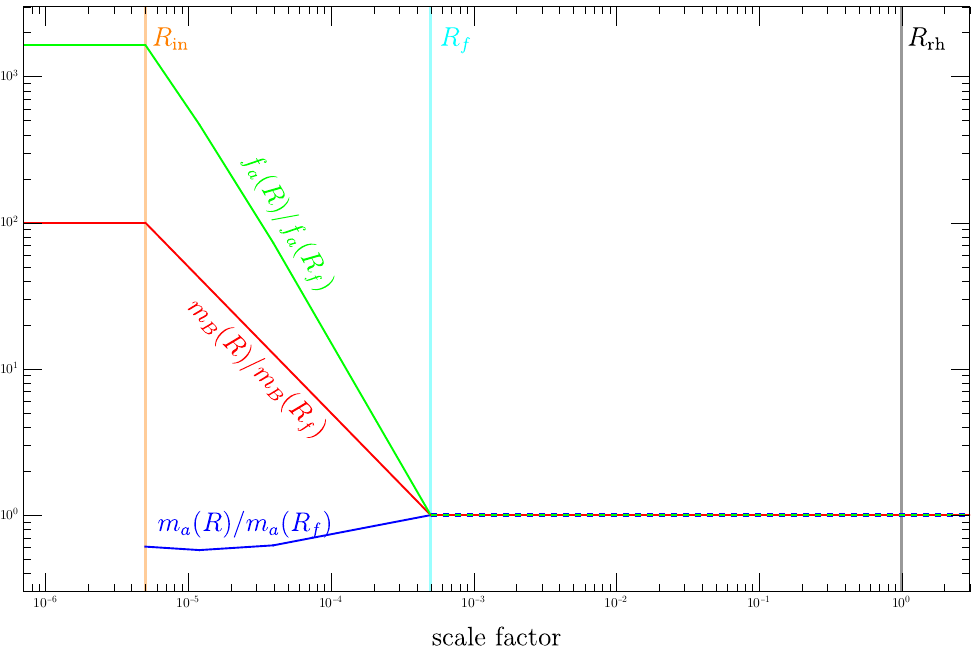}
    \caption{ 
    Plot of the axion decay constant ($f_a$), baryon mass ($m_B$), and axion mass $m_a$ as a function of scale factor, shown in green, red and blue colors respectively. 
    All these quantities are normalized to their values at the end of the relaxation, which are also their present-day values. 
    For $R\geq R_f$, all these quantities are equal to 1 in this normalization. This is shown by the solid-dashed line. The rest of the figure description matches that of  Fig.~\ref{fig:app_rho_various_R}. 
    }
    \label{fig:app_relaxation_R}
\end{figure}

In this section, we explicitly show how various quantities discussed in the main text vary as a function of the scale factor. 
For the purpose of plotting, we consider two decades of relaxation {\it i.e.} $R_{\rm in}/R_f=10^{-2}$ and take the final axion decay constant to be $f_a=10^9\GeV$. 
For these choices of parameters, the axion oscillation starts at $R=R_{\rm in}$. 
In all the figures in this section, the vertical orange, turquoise and the gray lines denote the scale factor associated to the beginning of relaxation ($R_{\rm in}$), end of the relaxation ($R_{f}$) and reheating ($R_{\rm rh}$) respectively.  
In Fig.~\ref{fig:app_rho_various_R}, we show the energy densities of the reheaton (green), baryon (blue), dark matter (red), and the relativistic part of the SM (black) as a function of scale factor. 
We see that the reheaton drives the expansion of the universe till $R=R_{\rm rh}$, and then it decays to the SM and reheats the universe at $\Trh\sim 10\MeV$. Then the standard cosmology begins. 
In Fig.~\ref{fig:app_SM_R}, we show the how do the SM temperature ($T_\gamma$) and baryon mass ($m_B$) change as a function of the scale factor. As discussed in the main text, $m_B\propto 1/R$ during the relaxation whereas it stays constant before and after it. 
On the other hand, as discussed in Sec.~\ref{sec:TSM}, the SM acquires temperature from the early decays of the reheaton field and the decay of the AD field. 
In Fig.~\ref{fig:app_SM_R}, we see that at early times the SM temperature is dominated by the decay of the AD field and scale as $T_\gamma \propto R^{-1}$, whereas close to the reheating, the reheaton decays become more important and thus the temperature decreases slowly, only as $T_\gamma \propto R^{-1/4}$. 
After the reheating, the universe is dominated by $\rho_\gamma$, and $T_\gamma \propto R^{-1}$ as expected.

In Fig.~\ref{fig:app_relaxation_R}, we show how the axion decay constant ($f_a$), baryon mass ($m_B$), and axion mass $m_a$ as a function of scale factor in green, red and blue colors respectively. 
We normalize these quantities to their respective values at the end of the relaxation which are also their present-day values. 
We see that during the relaxation phase, $f_a$ decreases faster with $R$ than $m_B\propto 1/R$. 
From Fig.~\ref{fig:app_relaxation_R}, we can roughly parameterize the $R$-dependence of $f_a$ as $f_a\propto R^{-3/2-\epsilon}$ where $|\epsilon|\ll 1$ is set by the IR QCD $\beta$-function appropriate to the number of light quarks compared to the instantaneous confinement scale. 
Note that, as the baryon mass and therefore the QCD confinement scale is changing during relaxation, $\epsilon$ is a function of the scale factor as well. We have thoroughly discussed this in the main text.  
As $m_a\propto \Lambda_{\rm QCD}^{3/2}/f_a\propto R^{\epsilon(R)}$, and $|\epsilon|\ll 1$, we see that the axion mass changes only slightly, by $\mathcal{O}(1)$ factor during the relaxation phase. As $\epsilon$ is a function of $R$, we also find that initially the axion mass decreases with the scale factor and then increases in two different rates towards the end of it.

\end{document}